\documentclass[aps,prd,notitlepage,amsmath,amssymb,preprintnumbers,nofootinbib,longbibliography,superscriptaddress,10pt]{revtex4-1}
\usepackage[utf8]{inputenc} 
\usepackage{graphicx}
\usepackage{url}
\usepackage[bookmarks, pagebackref=false]{hyperref}
\usepackage[usenames,dvipsnames]{xcolor}

\def\hc{\ensuremath{\mathrm{h.c.}}}

\def\BbsR{\ensuremath{\mathcal{B}^R_{bs}}}

\begin{document}

\title{\textbf{\huge Impact of a non-universal $Z^\prime$ on the $B\to K^{(*)}l^+l^-$ and $B \to K^{(*)}\nu\bar{\nu}$ processes}}

\author{A.V. {\sc Bednyakov}}\email{ bednya@jinr.ru}
\author{A.I. {\sc Mukhaeva}}\email{ mukhaeva@theor.jinr.ru}

\affiliation{%
	Joint Institute for Nuclear Research, Joliot-Curie, 6, Dubna 141980, Russia
}

\begin{abstract}
		We perform a study of the new physics effects in semileptonic FCNC processes 
		within a low-energy approximation of the anomaly-free supersymmetic extension of the SM with 
		additional $Z'$ vector field. The key feature of the model is the non-diagonal structure of $Z'$ 
		couplings to fermions, which is parameterized by few new-physics parameters in addition to well-known 
		mixing matrices for quarks and leptons in the SM. We not only consider CP-conserving scenarios with  
		real parameters, but also account for possible CP violation due to new physical weak phases. 
		We analyse the dependence of the $b\to s$ observables on the parameters together with 
		correlations between the observables predicted in the model.  
		Special attention is paid to possible enhancement of $B \to K^{(*)} \nu\bar{\nu}$ rates and 
		to CP-odd angular observables in $B \to K^* ll$ decays.
		
\end{abstract}
\maketitle
\section{Introduction\label{sec:intro}}

Flavor changing neutral current (FCNC) decays are expected to play a significant role in the search for physics beyond the Standard Model (SM) since they are loop-suppressed in the SM  and have enhanced sensitivity to the New Physics (NP) effect. Among interesting decays there are $b\to s$ transitions that have been the subject of attention due to the persistent observation of anomalies (see, e.g., \cite{Altmannshofer:2021qrr} and references therein), for example, in $B_s \to \mu^+ \mu^-$,  $B \to K^{{*}}l^+ l^-$, and also $B \to K^{(*)}\nu \bar{\nu}$ processes. These anomalies have been studied in two ways: a) by means of effective field theories (EFT) that include all possible new dimension-six operators, or b) building specific NP models. For the case of the EFT analyses, one performs global fits to all the $b\to s$ data with the aim of find the preferred Lorentz structure of the new-physics operators. 
As for specific Beyond-the-SM models (BSM), two main classes proposed to account for these anomalies are $Z^\prime$ models~\cite{Crivellin:2015mga,Crivellin:2015lwa,Allanach:2019mfl,Alok:2019xub,Allanach:2020kss,Allanach:2021gmj,Allanach:2022bik,Alok:2021pdh,Alok:2022pjb}, and models with leptoquarks (see, e.g., recent Ref.~\cite{London:2021lfn}).

Wilson coefficients (WC) in  NP scenarios can be real or complex, thereby giving rise to new sources of CP violation (CPV). Moreover, since CPV effects in $b\to s$ decays are suppressed in the SM \footnote{With the account of NLO QCD corrections and hadronic uncertainties, the CP asymmetries are still estimated to be $\lesssim 1\%$ \cite{Bobeth:2008ij},\cite{Altmannshofer:2008dz}} \cite{PhysRevD.61.114028}, \cite{PhysRevD.63.014013} these are promising channels to look for new sources of CP violation.  
The new CPV phases are very weakly constrained as there are only a few measurements of CPV observables. Global fits with real (complex)  NP WCs have been performed in Refs.\cite{Alguero:2021anc}(\cite{Altmannshofer:2021qrr,Alok:2017jgr}), and a few studies in the past obtained constraints on the parameter space of $Z^\prime$ and leptoquark models \cite{Alok:2017jgr,DiLuzio:2019jyq,Alok:2019xub}.  The most relevant CPV observables for our analysis are 
direct CP asymmetries in $B\to K^{(*)}\mu^+\mu^-$, and CP asymmetric angular observables $A_7$, $A_8$ and $A_9$ measured by LHCb \cite{LHCb:2014mit,LHCb:2015svh}, which still have large uncertainties and are consistent with zero.

The main goal of this paper is to extend the study of Ref.~\cite{Bednyakov:2021fof}
of 
simplified scenario with heavy $Z'$ boson possessing FCNC coupling at the tree level. 
In the context of this model we want to explain deviations from the SM predictions in the neutral current (NC) channels such as  a set of angular observables for 
$B \to K^* \mu^+\mu^-$; the branching ratio $B_s \to \mu^+ \mu^-$; and the $B_s$ mixing data. 
We also take into account updated results from LHCb collaboration \cite{LHCb:2022zom} on the lepton flavor universality ratios $R_K$ and $R_{K^*}$, which turned out to be compatible with the SM.

In addition, given unique Belle II (see, e.g., \cite{Belle:2019oag,Bose:2022obr}) capabilities to measure $B \to K^{(*)}\nu \bar{\nu}$  branching ratios, we want to check possible enhancement  in these processes. There have been previous studies  analyzing the effect of NP models in $B \to K^{(*)}\nu \bar{\nu}$  modes , some focusing on the connection with $b \to s \mu^+\mu^-$ anomalies in an effective theory approach (see, e.g. Refs.~\cite{Descotes-Genon:2020buf,Rajeev:2021ntt}). However, in this work we explore the NP parameter space connecting  $b \to s \mu^+\mu^-$  tensions and also include relatively light right-handed neutrinos (RHNs). Thus, the computations of these NP contributions are performed in our model.

The predicted non-diagonal flavour structure of the $Z'$ couplings are related to CKM and PMNS matrices and is parametrized by additional mixing angles and complex phases (see Ref.~\cite{Bednyakov:2021fof} for more detail).
These complex phases propagate to the low-energy effective Hamiltonian and can account for new CP-violating effects in the angular observables of $B \to K^{{*}}l^+ l^-$. We carry out comprehensive analysis of the model constrained by available experimental data, and study possible CPV manifestation in these decays.

The paper is organized as follows: Starting with general dimension-6 effective Hamiltonian (including RHNs), in Sec.~\ref{sec:weak_hamiltonian}, we consider $U(1)^\prime$ extension of MSSM with additional $Z^\prime$-boson in Sec.~\ref{sec:UnuMSSM}.  Then we
review  observables  for semileptonic $B\to K^{(*)}$ transitions in Sec.~\ref{sec:BKll_obs}. 
Discussion of the fit procedure together with our phenomenological analysis can be found in Sec.~\ref{sec:fit}.  We conclude in Sec.~\ref{sec:res}.

\section{Weak effective Hamiltonian for $b\to s $ transition\label{sec:weak_hamiltonian}}

The general dimension-6 effective Hamiltonian relevant for $b\to s ll
$  and $b\to s \nu \bar{\nu}$  transitions including light RHN fields can be written as
\begin{align}
     \mathcal{H}_{eff}^{b\to s ll} = -\frac{4G_F}{\sqrt{2}}\frac{\alpha_{e}}{4\pi}V_{tb} V_{ts}^* \left[ C_{L}^{SM}\delta_{\alpha\beta}O_{L}^{\alpha\beta} + \sum_{\alpha\beta}\left(\sum_{i=L^{(')},R^{(')}} C_{i}^{\alpha\beta} O_{i}^{\alpha\beta} + \sum_{j=  9^{(')},10^{(')}}O_j^{\alpha\beta} C_j^{\alpha\beta}\right) \right]  + \mathrm{h.c.},
   \label{eq:Heff}
\end{align}
and the NP contribution to $B_s-\bar{B}_s$ mixing can be parameterized by the effective Hamiltonian
\begin{align}
     \mathcal{H}_{eff}^{\Delta B=2} = -\frac{4G_F}{\sqrt{2}}(V_{tb} V_{ts}^*)^2  \sum_{i= LL,LR,RR} C_i^{bs}O_i^{bs}  + \mathrm{h.c.},
   \label{eq:Heff_Ms}
\end{align}
where $G_F$ is the Fermi constant and $V_{ij}$ denote the Cabibbo–Kobayashi–Maskawa (CKM) matrix elements.  The short distance contributions are encoded in the Wilson coefficients $C_i$ of the four-fermi operators $O_i$. The scale dependence is implicit here, $C_i \equiv C_i(\mu)$ and $O_i \equiv O_i(\mu)$, and, if not stated otherwise, we choose the scale to be around the bottom-quark mass $\mu\simeq m_b$.

Four-fermion operators given in \eqref{eq:Heff} have the following form:
\begin{align}
    O_{L}^{\alpha\beta} = (\bar{s}_L\gamma^\mu  b_L)(\bar{\nu}^\alpha \gamma_\mu (1-\gamma_5) \nu^\beta),\qquad O_{R}^{\alpha\beta} = (\bar{s}_R\gamma^\mu  b_R)(\bar{\nu}^\alpha \gamma_\mu (1-\gamma_5) \nu^\beta),\nonumber\\
    O_{L}^{'\alpha\beta} = (\bar{s}_L\gamma^\mu  b_L)(\bar{\nu}^\alpha \gamma_\mu (1+\gamma_5) \nu^\beta), \qquad O_{R}^{'\alpha\beta} = (\bar{s}_R\gamma^\mu  b_R)(\bar{\nu}^\alpha \gamma_\mu (1+\gamma_5) \nu^\beta),\nonumber\\
    O_9^{\alpha\beta} = (\bar{s}_L\gamma^\mu  b_L)(\bar{l}^\alpha \gamma_\mu l^\beta),\qquad  O_{10}^{\alpha\beta} = (\bar{s}_L\gamma^\mu  b_L)(\bar{l}^\alpha \gamma_\mu \gamma_5 l^\beta), \nonumber\\
     O_9^{'\alpha\beta} = (\bar{s}_R\gamma^\mu  b_R)(\bar{l}^\alpha \gamma_\mu l^\beta),\qquad  O_{10}^{\alpha\beta} = (\bar{s}_R\gamma^\mu  b_R)(\bar{l}^\alpha \gamma_\mu \gamma_5 l^\beta),\nonumber\\
     O_{LL(RR)}^{bs} = (\bar{s}_{L(R)}\gamma^\mu  b_{L(R)})(\bar{s}_{L(R)}\gamma^\mu  b_{L(R)}),\qquad O_{LR}^{bs} = (\bar{s}_{L}\gamma^\mu  b_{L})(\bar{s}_{R}\gamma^\mu  b_{R}),
\end{align}
and are not necessary diagonal in the lepton flavor indices $\alpha,\beta$.

We exclude effective operators with scalar and tensor neutrino bilinears\footnote{
Both of them were considered in Ref.~\cite{Browder:2021hbl}.} 
	from Eq.~\eqref{eq:Heff} since they do not appear from the tree-level $Z'$ exchange in our model.

The  SM contribution to $C_9$ and $C_{10}$ at the scale $\mu = m_b = 4.8$ GeV, to NNLL accuracy \cite{Altmannshofer:2008dz} is given by:
\begin{align}
    C_9^{SM} = 4.211, \qquad C_{10}^{SM} = -4.103.
    \label{eq:C9_C10_SM}
\end{align}
For the operators with neutral leptons, the SM gives rise to 
the diagonal Wilson coefficient \cite{Descotes-Genon:2020buf}:
\begin{align}
	C_L^{\alpha\alpha} \equiv C_{L}^{SM} =-2X_t/s_w^2  
    \label{eq:CL_SM}
\end{align}
with $X_t = 1.469\pm 0.017$, includes NLO QCD corrections and two-loop electroweak contributions. All other Wilson coefficients $C_{i}^{\alpha\beta} = 0$ (except for a negligible contribution to $O_{R}^{\alpha\alpha}$) in the SM, and thus any nonzero contribution to these Wilson coefficients, is then a manifestation of NP beyond the SM. 

Turning to $B_s-\bar{B}_s$ mixing, the SM contribution arises due to a box diagram, and is given by
\begin{align}
    C_{LL}^{bs(SM)} = \eta_{B_s}x_t\left[ 1+ \frac{9}{1-x_t} - \frac{6}{(1-x_t)^2 } - \frac{6x_t^2 \ln{x_t}}{(1-x_t)^3}\right],
    \label{eq:CLL_SM}
\end{align}
here $x_t \equiv m_t^2/m_W^2 $ and $\eta_{B_s} = 0.551$  is the QCD correction \cite{Buchalla:1995vs}. 

\section{Low-energy limit of the $U\nu_R MSSM$ model \label{sec:UnuMSSM}}
We consider a non-universal $Z^\prime$ effective model, where heavy $Z^\prime$ boson is associated with an additional non-anomalous $U(1)^\prime$ symmetry in the non-minimal supersymmetric extension of the SM proposed in Ref.~\cite{Bednyakov:2021fof}. 
The $Z^\prime$  boson couples to both left and right-handed leptons. Further, the couplings to both left and right-handed quarks are allowed
\footnote{We neglect the mixing with the SM $Z$ boson.} 
\begin{align}
    \Delta \mathcal{L}_{Z^\prime} = g_E J^\alpha Z_\alpha^\prime.
\end{align}
Here $g_E$ is the $U(1)^\prime$ gauge coupling, and the fermionic current is given in terms of up ($\mathcal{U}_{q}$) and down ($\mathcal{D}_{q}$) quarks,  charged ($\mathcal{E}_l$) and neutral ($\mathcal{N}_\nu$) leptons 
(see Ref.~\cite{Bednyakov:2021fof}). The current includes
\begin{align}
J^\alpha  \supset &
 \sum\limits_{q,q^\prime=1,3}
\left[
V_{R,3q}V^*_{R,3q^\prime} (\overline{\mathcal{D}}_{qR} \gamma_\alpha
\mathcal{D}_{q^\prime R})
+ V_{L,3q}V^*_{L,3q^\prime} (\overline{\mathcal{D}}_{qL} \gamma_\alpha
\mathcal{D}_{q^\prime L})
\right] \nonumber\\
&
-\sum\limits_{l=1,3}
\left[
\overline{\mathcal{E}}_{l} \gamma_\alpha
\mathcal{E}_{l^\prime}
+ \overline{\mathcal{N}}_{l} \gamma_\alpha
\mathcal{N}_{l^\prime}
- 
V^*_{L,3l} V_{L,3l^\prime}
(\overline{\mathcal{E}}_{l L} \gamma_\alpha
\mathcal{E}_{l^\prime L})
\right]\nonumber\\
& 
+\sum\limits_{\nu\nu^\prime=1,3}
\left[
V^*_{ L,3\nu}V_{ L,3\nu^\prime}
(\overline{\mathcal{N}}_{\nu L} \gamma_\alpha
\mathcal{N}_{\nu^\prime L})
+
V^*_{R,3\nu} V_{ R,3\nu^\prime} (\overline{\mathcal{N}}_{\nu R} \gamma_\alpha
\mathcal{N}_{\nu^\prime R})
\right].
\label{eq:lag_Z_mass_eigenstates}
\end{align}
In Eq.~\eqref{eq:lag_Z_mass_eigenstates} the mixing-matrices elements for quarks $V_{L(R), 3q}$ are defined as 
\begin{align}
V_{L,3q} & = \left\{  - s^d_{13}e^{-i \phi_{13}} , -c^d_{13} s^d_{23}e^{-i\phi_{23}} ,  c^d_{13} c^d_{23} \right\}, \nonumber \\
V_{R,3q} & = \frac{\left\{
	-m_b m_s s^d_{13} e^{-i\phi_{13}},
	-m_b m_d c^d_{13} s^d_{23} e^{-i \phi_{23}},
	m_s m_d c^d_{13} c^d_{23}
	\right\}
}{
	\sqrt{m_d^2 (m_b^2 s_{23}^{2} + m_s^2 c_{23}^{2}) c_{13}^{2} + m_b^2 m_s^2 s_{13}^{2}}
},		
\label{eq:quark_Vi3}
\end{align}
while for leptons one can write
\begin{align}
V_{L,3l} & = \left\{
- s^e_{13}e^{i \chi_{13}} ,
- c^e_{13} s^e_{23}e^{i \chi_{23}} ,
c^e_{13} c^e_{23}
\right\}, \qquad V_{R,3l} = 1, \\
V_{L,3\nu} & =
\left\{
\tilde U_{l1} , \tilde U_{l2} , \tilde U_{l3}
\right\},
\quad 
V_{R,3\nu}  =
\frac{
	\left\{
	m_{\nu_1} \tilde U_{l1},
	m_{\nu_2} \tilde U_{l2},
	m_{\nu_3} \tilde U_{l3}
	\right\} }
{\sqrt{m^2_{\nu_3} |\tilde U_{l3}|^2 + m^2_{\nu_2} |\tilde U_{l2}|^2 +m^2_{\nu_1} |\tilde U_{l1}|^2 }}.
\label{eq:lepton_V3i}
\end{align}
For convenience, we introduce the following shorthand notation
\begin{align}
\tilde U_{li} & \equiv c^e_{13} (U_{\tau i} c_{23}^e - U_{\mu i} s_{23}^e e^{-i\chi_{23}})
- U_{ei} s_{13}^e e^{-i \chi_{13}}, \qquad i = \{1,2,3\},
\label{eq:tildeU_tildeV}
\end{align}
with $U_{l_i j}$ being the matrix elements of PMNS matrix. 

The mixing matrices ~\eqref{eq:quark_Vi3}-\eqref{eq:lepton_V3i} incorporate new model parameters as angles and phases, where $c_{13, 23}^{d,e}$, $s_{13, 23}^{d,e}$  -- mixing angles between  1 and 3, 2 and 3 generation of quarks and leptons, respectively, with $c_\alpha \equiv \cos\alpha$, $s_\alpha \equiv \sin\alpha$, and $\phi_{13, 23}$, $\chi_{13,23}$ -- new CP-violating phases  of quarks and leptons.

We can introduce the following notation
\begin{align}
   g_{L}^{qq^\prime} \equiv V_{L,3q}V^*_{ L,3q^\prime},  \qquad g_{R}^{qq^\prime} \equiv V_{R,3q}V^*_{ R,3q^\prime}, \nonumber\\
   g_{L}^{ll^\prime} \equiv V_{L,3l}V^*_{ L,3l^\prime} - \delta_{ll^\prime},  \qquad g_{R}^{ll^\prime} \equiv 1, \nonumber\\
   g_{L}^{\nu\nu^\prime} \equiv V_{ L,3\nu}V^*_{ L,3\nu^\prime} - \delta_{\nu\nu^\prime},  \qquad g_{R}^{\nu\nu^\prime} \equiv V_{ R,3\nu}V^*_{ R,3\nu^\prime} - \delta_{\nu\nu^\prime},
\end{align}
where $g_{L(R)}^{ll^\prime}$ are the left-handed (right-handed) couplings of the $Z^\prime$ boson to leptons, $g_{L(R)}^{\nu\nu^\prime}$ to neutrinos and $g_{L(R)}^{qq^\prime}$ to quarks.

It is worth noting that the model predicts right-handed neutrinos, which are SM singlets and have Dirac-type masses $m_{\nu_i}$. In this study, we consider  quasi-degenerate case (see, e.g., Ref.~\cite{ParticleDataGroup:2022pth}) corresponding to $m_{\nu_1} \simeq m_{\nu_2}  \simeq m_{\nu_3}  \lesssim 0.1 $~eV, i.e., all the masses are  greater than the mass difference, but negligible compared to $q^2$ considered in the $B\to K^{(*)}\nu\bar{\nu}$ transitions. In this case we have $V_{\nu L} = V_{\nu R}$, and $Z'$ ceases to couple with the neutrino axial current.

After integrating out the heavy  $Z^\prime$, we get the effective four-fermion Hamiltonian. 
The relevant terms in the effective Hamiltonian is given by
\begin{align}
	\mathcal{H}_{eff}^{Z^\prime} = \frac{ g_E^2}{2M_{Z^\prime}^2}J_\alpha J^\alpha & \supset \frac{ g_E^2}{M_{Z^\prime}^2} g_{L}^{bs}(\bar{s}\gamma^\alpha P_L b)[\bar{l}\gamma_\alpha (g^{ll'}_{L} P_L + g^{ll'}_{R} P_R)l'] \nonumber\\
										       & + \frac{ g_E^2}{M_{Z^\prime}^2} g_{R}^{bs}(\bar{s}\gamma^\alpha P_R b)[\bar{l}\gamma_\alpha (g^{ll'}_{L} P_L + g^{ll'}_{R} P_R) l] \nonumber\\
                                              & + \frac{ g_E^2}{2M_{Z^\prime}^2} (g_{L(R)}^{bs})^2(\bar{s}\gamma^\alpha P_{L(R)} b)(\bar{s}\gamma^\alpha P_{L(R)} b) \nonumber\\
					      & + \frac{ g_E^2}{M_{Z^\prime}^2} (g_{L}^{bs})(g_{R}^{bs})(\bar{s}\gamma^\alpha P_L b)(\bar{s}\gamma^\alpha P_R b) \nonumber\\
										       & + \frac{ g_E^2}{M_{Z^\prime}^2}g_{L}^{bs}(\bar{s}\gamma^\alpha P_L b)[\bar{\nu}\gamma_\alpha (g^{\nu\nu'}_{L} P_L + g^{\nu\nu'}_{R} P_R)\nu'] \nonumber\\
										       & + \frac{ g_E^2}{M_{Z^\prime}^2} g_{R}^{bs}(\bar{s}\gamma^\alpha P_R b)[\bar{\nu}\gamma_\alpha (g^{\nu\nu'}_L P_L + g^{\nu\nu'}_R P_R)\nu'] + \hc
    \label{eq:Heff_Zprime}
\end{align}
Here  $M_{Z'}$ denotes the $Z'$-boson mass. 
Comparing Eq.~\eqref{eq:Heff_Zprime} with Eq.~\eqref{eq:Heff}, one gets the expressions for the Wilson coefficients induced by the $Z'$ exchange 
\begin{align}
C_9^{ll^\prime} = \mathcal{N} \frac{ g_E^2}{M_{Z^\prime}^2} g_L^{bs}[g_R + g_L]^{ll^\prime} \qquad C_9^{\prime ll^\prime} = \mathcal{N} \frac{ g_E^2}{M_{Z^\prime}^2} g_R^{bs}[g_R + g_L]^{ll^\prime},
\label{eq:C9}\\
C_{10}^{ll^\prime}= \mathcal{N} \frac{ g_E^2}{M_{Z^\prime}^2} g_L^{bs}[g_R - g_L]^{ll^\prime} \qquad C_{10}^{\prime ll^\prime} = \mathcal{N}\frac{ g_E^2}{M_{Z^\prime}^2} g_R^{bs}[g_R - g_L]^{ll^\prime},
\label{eq:C10}\\
C_L^{\nu\nu^\prime} = \mathcal{N}\frac{ g_E^2}{M_{Z^\prime}^2} g_L^{bs}[g_L]^{\nu\nu^\prime} \qquad C_L^{\prime \nu\nu^\prime} = \mathcal{N} \frac{ g_E^2}{M_{Z^\prime}^2} g_L^{bs}[g_R]^{\nu\nu^\prime},
\label{eq:CL}\\
C_R^{\nu\nu^\prime} = \mathcal{N}\frac{ g_E^2}{M_{Z^\prime}^2}g_R^{bs} [g_L]^{\nu\nu^\prime} \qquad C_R^{\prime \nu\nu^\prime} = \mathcal{N}\frac{ g_E^2}{M_{Z^\prime}^2} g_R^{bs}[g_R]^{\nu\nu^\prime},
\label{eq:CR}
\end{align}
\begin{align}
    C_{LL(RR)}^{bs} = - \frac{1}{4\sqrt{2}G_F (V_{tb}V_{ts}^*)^2} \frac{ g_E^2}{M_{Z^\prime}^2} (g_{L(R)}^{bs})^2 \qquad C_{LR}^{bs} = - \frac{1}{2\sqrt{2}G_F (V_{tb}V_{ts}^*)^2} \frac{ g_E^2}{M_{Z^\prime}^2} (g_L^{bs})(g_R^{bs}),
\label{eq:CLL}
\end{align}
where the overall factor is given by $\mathcal{N} = -\frac{\pi}{\sqrt{2}G_F\alpha_{e} V_{tb}V_{ts}^*}$.

\section{Observables for semileptonic $B \to K^{(*)}$ transitions \label{sec:BKll_obs}}

In this section we briefly review the key observables used in our phenomenological analysis.
\subsection{Decay into charged leptons \label{sec:BKll}}

The differential distribution of $\bar{B}^0 \to \bar{K}^{*0}(\to K^- \pi^+) l^+l^-$ decay can be parametrized in terms of one kinematic and three angular variables. The kinematic variable is the invariant mass of the lepton pair, $q^2 = (p_B - p_{K^*})^2 = (p_{l^+}+p_{l^-})^2$, where $p_B$, $p_{K^*}$, and $p_{l^\pm}$the four-momenta of $\bar{B}$, $K^*$ mesons, and charged leptons, respectively. There are several conventions to define angular variables (see, e.g., Ref.~\cite{Gratrex:2015hna}).  
We consider 1)  the angle $\theta_K$ of $K^-$ in the rest frame of $\bar{K}^{*}$ with respect to the direction of flight of the latter in the $\bar{B}$ rest system;
2) the angle  $\theta_l$ of $l^-$ in the dilepton rest frame with respect to the direction of the lepton pair in the $\bar{B}$ rest frame;
3) the angle $\phi$ between $K^-\pi^+$ decay plane and the plane defined by the dilepton momenta.

The full angular decay distribution of $\bar{B}^0 \to \bar{K}^{*0}(\to K^- \pi^+) l^+l^-$ \cite{Bobeth:2008ij} can be cast into the form
\begin{equation}
\label{eq:diffdec}
\frac{d^4\Gamma}{dq^2 d\cos\theta_{l}d\cos\theta_{K}d\phi} = \frac{9}{32\pi}J(q^2, \theta_{l}, \theta_{K}, \phi),
\end{equation}
where
\begin{align}
\label{eq:J}
 J(q^2, \theta_{l}, \theta_{K}, \phi) = {}\nonumber\\ 
    J_{1s}\sin^{2}\theta_{K} + J_{1c}\cos^{2}\theta_{K} + (J_{2s}\sin^{2}\theta_{K} + J_{2c}\cos^{2}\theta_{K})\cos 2\theta_{l} +  {}\nonumber\\
    J_{3}\sin^{2}\theta_{K}\sin^{2}\theta_{l}\cos 2\phi + J_{4}\sin 2\theta_{K}\sin 2\theta_{l}\cos \phi + {}\nonumber\\ 
    J_{5}\sin 2\theta_{K}\sin\theta_{l}\cos \phi + (J_{6s}\sin^{2}\theta_{K} + J_{6c}\cos^{2}\theta_{K})\cos \theta_{l} + {}\nonumber\\
    J_{7}\sin 2\theta_{K}\sin \theta_{l}\sin \phi +  J_{8}\sin 2\theta_{K}\sin 2\theta_{l}\sin \phi +  {}\nonumber\\
    J_{9}\sin^{2}\theta_{K}\sin^{2}\theta_{l}\sin 2\phi.
\end{align}
The expressions of these twelve angular coefficients $J_i(a) $ are well known from literature (see, e.g., Ref.~\cite{Altmannshofer:2008dz}). 
These coefficients depend on the $q^2$ variable, on Wilson coefficients and various hadronic form factors. 
The corresponding expression for the four-fold decay distribution of the CP conjugate decay mode $B^0 \to K^{*0}(\to K^+ \pi^-) l^-l^+$ can be obtained by substituting $\theta_l$ by $(\pi -\theta_l)$ and $\phi$ by $-\phi$. This results in the following transformations of angular coefficients
\begin{align}
    J_{1,2,3,4,7}^{(a)} \to \bar{J}_{1,2,3,4,7}^{(a)}, \qquad J_{5,6,8,9}^{(a)} \to -\bar{J}_{5,6,8,9}^{(a)}.
\end{align}
Here $\bar{J}_{i}^{(a)}$ equal to $J_i^{(a)}$, in which all \emph{weak} phases are conjugated.

The angular coefficients have a clear relation to both experiment and theory: theoretically they are expressed in terms of transversity amplitudes, and experimentally they describe the angular distribution.  For example,  $J_7$ to $J_9$ depend on the imaginary part of the transversity amplitudes, and consequently on their phases, which come either from QCD effects and enter the QCD factorization expressions at $O(\alpha_s)$, or are CP-violating SM or NP phases. To separate CP-conserving and CP-violating NP effects, it is more convenient to consider the twelve CP averaged angular coefficients \cite{Bobeth:2008ij}
\begin{align}
    S_i^{(a)}(q^2) = \frac{J_i^{(a)}(q^2)+\bar{J}_{i}^{(a)}(q^2)}{d(\Gamma + \bar{\Gamma})/dq^2}, 
\end{align}
as well as the twelve CP asymmetries
\begin{align}
    A_i^{(a)}(q^2) = \frac{J_i^{(a)}(q^2)-\bar{J}_{i}^{(a)}(q^2)}{d(\Gamma + \bar{\Gamma})/dq^2}.
\end{align}
The CP asymmetry in the dilepton mass spectrum is defined as
\begin{align}
    A_{CP}(q^2) = \frac{d\Gamma/dq^2-d\bar{\Gamma}/dq^2}{d\Gamma/dq^2+d\bar{\Gamma}/dq^2},
\end{align}
where $d\Gamma/dq^2$ can be expressed in terms of angular coefficients as
\begin{align}
    \frac{d\Gamma}{dq^2} = \frac{3}{4}(2J_1^s+J_1^c)-\frac{1}{4}(2J_2^s+J_2^c).
\end{align}

In what follows we consider the impact of new complex phases on the angular distributions of the CP-conjugated decay modes.   
One can distinguish two types of CPV effects: the direct CP violating asymmetries and triple-product CP asymmetries. 
Let us consider two amplitudes $A_1 \propto e^{i\phi_1}e^{i\delta_1}$ and $A_2 \propto e^{i\phi_2}e^{i\delta}$ contributing to the $b\to s l^+l^-$ process. 
Here $\phi_{1,2}$ and $\delta_{1,2}$  are weak and strong phases, respectively. It can be shown that the direct CP asymmetries are proportional to $\mathrm{sin}(\phi_1-\phi_2) \mathrm{sin}(\delta_1-\delta_2)$. This means that the asymmetries can have non-zero values only if the two interfering amplitudes have a relative weak and a strong phase. On the contrary, triple-product asymmetries are proportional to $\mathrm{sin}(\phi_1-\phi_2) \mathrm{cos}(\delta_1-\delta_2)$. As a consequence, it is sufficient to have only a relative weak phase between the amplitudes to provide a non-zero value. 
The SM has a finite strong phase emanating from the imaginary contribution to $C_9^{eff}$, which is generated by the $q\bar{q}$ loops in the current-current quark operators.
However, the weak phase, coming from the CKM elements, is double Cabibbo-suppressed and small. Therefore, the CP violation in the SM is not expected to be large.

In this work we consider both direct ($A_{3,4,5,6s}$, $A_{CP}$) and triple-product ($A_{7,8,9}$) CP asymmetries.  These observables are measured by the LHCb collaboration, however, with large errors \cite{LHCb:2015svh}. Observation of non-zero CP asymmetries in $b\to sll$ decays would be a clear signature of new physics. In the absence of a non-zero signal, precise measurements of the CP asymmetries $A_{7,8,9}$  can provide important bounds on BSM sources of CP violation in the form of imaginary parts of the 
Wilson coefficients. 

\subsection{Decay into neutrinos \label{sec:BKnunu}}
We also consider processes with neutral leptons in the final state. First, we focus on the differential decay distributions for $B\to (P,V)\nu \bar{\nu}$  where $P$ and $V$ denote the pseudoscalar and vector mesons, respectively. For the $B\to P\nu \bar{\nu}$ case we have \cite{Altmannshofer:2009ma}
\begin{align}
    \frac{d\Gamma(B\to P\nu \bar{\nu})}{dq^2} = \frac{4 G_F^2 \alpha^2}{256\pi^5 m_B^3} |V_{tb}V_{ts}^*|^2 \lambda^{3/2}(m_B^2,m_P^2,q^2)[f_+(q^2)]^2 
    \sum_{\alpha=1}^3\sum_{\beta=1}^3 \left[|C_V^{\alpha\beta}|^2 + |C_V^{\prime\alpha\beta}|^2 \right],
    \label{eq:BP}
\end{align}
where $2 C^{(')}_{V,A} = C^{(')}_R \pm C^{(')}_L$ are analogs of $C^{(')}_{9,10}$ in the charged lepton case.
The function $\lambda$ has a usual definition $\lambda(a,b,c) = a^2+b^2+c^2-2(ab+bc+ca)$, and the form factor $f_+(q^2)$ is taken from lattice QCD computations \cite{Bailey:2015dka}.

Similarly, for the decay into vector meson $B\to V\nu \bar{\nu}$ 
\begin{align}
\frac{d\Gamma(B\to V\nu \bar{\nu})}{dq^2} = 4 \sum_{\alpha=1}^3\sum_{\beta=1}^3  
\left(
\left[ A_0^2 + A_{\parallel}^2\right] \left[|C_A^{\alpha\beta}|^2 + |C_A^{\prime\alpha\beta}|^2\right] 
+ A_{\perp}^2 \left[|C_V^{\alpha\beta}|^2 + |C_V^{\prime\alpha\beta}|^2\right].
\right)
\label{eq:BV}
\end{align}
The quantities $A_0$, $A_\perp$, $A_\parallel$ are the $B\to V$ transversity amplitudes, which are given by 
\begin{align}
    A_{\perp}(q^2) = \frac{2 \mathcal{M} \sqrt{2\lambda(m_B^2,m_V^2,q^2)}}{m_B^2}\frac{V(q^2)}{\left[1+\frac{m_V}{m_B} \right]},\\
    A_{\parallel}(q^2) = -2\mathcal{M}\sqrt{2}\left[1+\frac{m_V}{m_B} \right]A_1(q^2),\\
    A_0(q^2) = -\frac{\mathcal{M} m_B^2}{m_V\sqrt{q^2}}\left(
    \left[1-\frac{m_V^2}{m_B^2} - \frac{q^2}{m_B^2} \right]\left[1+\frac{m_V}{m_B} \right]A_1(q^2)\right. \nonumber\\\left. - \frac{\lambda(m_B^2,m_V^2,q^2)}{m_B^4}\frac{A_2(q^2)}{\left[1+\frac{m_V}{m_B} \right]}\right) \equiv -\frac{16 \mathcal{M} m_V}{\sqrt{q^2}} A_{12}(q^2),
\end{align}
where $A_{12}(q^2) = \frac{(m_B + m_V)^2 (m_B^2 - m_V^2 - q^2)A_1 - \lambda A_2}{16 m_B m_V^2 (m_B + m_V)}$. 
Here  $\mathcal{M}$ is the normalization factor with $q^2$ being the invariant mass of the neutrino-antineutrino pair
\begin{align}
    \mathcal{M} = |V_{tb}V_{ts}^*| \left[ \frac{G_F^2 \alpha^2 q^2 \sqrt{\lambda(m_B^2,m_V^2,q^2)}}{3 \times 2^{10}\pi^5 m_B} \right]^{1/2}.
\end{align}
The handronic form factors $V(q^2)$, $A_1(q^2)$, $A_2(q^2)$ are from a combined LCSR and lattice QCD analysis \cite{Bharucha:2015bzk}.

In addition to the differential decay distribution, in the case of vector meson in the final state one defines a longitudinal polarization fraction $F_L$, which can be written as
\begin{align}
    F_L = \frac{ 4 |A_0|^2}{d\Gamma/dq^2} \sum_{\alpha=1}^3\sum_{\beta=1}^3 \left(|C_A^{\alpha\beta}|^2+|C_A^{\prime\alpha\beta}|^2\right) .
    \label{eq:FL_H}
\end{align}
The normalization of $F_L$ on the total dineutrino spectrum significantly reduces the hadronic uncertainties associated with the form factors as well as the uncertainties associated with CKM elements.

In what follows we only study the quantities integrated over the whole available kinematic region, i.e., branching ratios for $B\to K^{(*)}\nu\bar{\nu}$ and $\langle F_L \rangle$. In the latter observable one independently integrates (or averages) the numerator and the denominator of \eqref{eq:FL_H} over $q^2$. 

To facilitate the comparison with the SM, we present our predictions for the ratios $R^{\nu\bar{\nu}}_P$, $R^{\nu\bar{\nu}}_V$ and $R^{\nu\bar{\nu}}_{F_L}$ where, $P$ or $V$ represent pseudoscalar or vector mesons 
\cite{Buras:2014fpa}:
\begin{align}
	R^{\nu\bar{\nu}}_P = \frac{\Gamma(B \to P \nu\bar{\nu})_{\phantom{SM}}}{\Gamma(B \to P \nu\bar{\nu})_{SM}}, 
	\qquad R^{\nu\bar{\nu}}_V = \frac{\Gamma(B \to V \nu\bar{\nu})_{\phantom{SM}}}{\Gamma(B \to V \nu\bar{\nu})_{SM}}, 
	\qquad R^{\nu\bar{\nu}}_{F_L} =\frac{\langle F_L \rangle (B \to V \nu\bar{\nu})_{\phantom{SM}}}{\langle F_L \rangle (B \to V \nu\bar{\nu})_{SM}}. 
	\label{eq:Rnunu}
\end{align}

In our scenario with quasi-degenerate neutrinos all NP contributions to $C_A$ and $C'_A$ are zero, so the numerator in Eq.~\eqref{eq:FL_H} is not modified w.r.t the SM. However, the denominator \eqref{eq:BV} can be affected by NP, and we see that for $R^{\nu\bar{\nu}}_V\geq1$, we should have $R^{\nu\bar{\nu}}_{F_L}\leq1 $, and vice versa.   
To compute the relevant observables we used a modified version of \texttt{flavio} \cite{Straub:2018kue} and \texttt{wilson} \cite{Aebischer:2018bkb} packages to account for right-handed neutrinos (see Ref.~\cite{Bednyakov:2022}).

\section{Fit results and model predictions \label{sec:fit}}

Our statistical analysis is based on the likelihood function defined for the set of NP input parameters $m$ given in Sec.~\ref{sec:UnuMSSM} as 
\begin{align}
    \mathcal{L}(m) = \exp \Bigg[-\frac{1}{2}[\mathcal{O}^{th}(m)-\mathcal{O}^{exp}]^T (\mathcal{C}^{exp}+\mathcal{C}^{th})^{-1}[\mathcal{O}^{th}(m)-\mathcal{O}^{exp}]
    \Bigg].
\end{align}
Here $\mathcal{O}^{th}(m)$ are the theoretical predictions of the observables calculated using  \texttt{flavio}, and $\mathcal{O}^{exp}$  are the corresponding experimental measurements.
The matrix $\mathcal{C}^{exp}$ encodes experimental correlation. The experimental correlation is available in angular observables for $B\to K^* \mu^+\mu^-$ \cite{LHCb:2015svh} and $B_s \to \phi \mu^+\mu^-$ \cite{LHCb:2015wdu}. For the other observables, we add the statistical and systematic errors in quadrature. If the errors are asymmetric, we use the larger error on both sides of the central value.

The theoretical correlation is given by the matrix $\mathcal{C}^{th}$ computed using the \texttt{flavio} \cite{Straub:2018kue} package, where hadronic form factors from lattice QCD are implemented.  The theoretical uncertainties are estimated as the standard deviation of the values of the observables, calculated by taking $N$ random choices of all input parameters according to their probability distribution. In this procedure we take $N = 2000$ random points, which corresponds to a $\sim 2\%$ precision on the theoretical error estimate. 

We performed two type of fits: with and without CPV observables. The statistical analysis performed in this study takes into account a large set of experimental measurements involving $b \to s$ transitions as implemented in \texttt{flavio}. In the following we summarize them briefly: (i) $B_s \to \mu^+\mu^-$ branching ratio; (ii)  $R_K$; (iii) $R_K^*$  for $B^0$ as well as $B^+$ decays; (iv) the differential branching ratios of $B_d \to K^*\mu^+\mu^-$, $B^+ \to K^* \mu^+\mu^-$, $B_d \to K \mu^+\mu^-$, $B^+\to K^+\mu^+\mu^-$ and $B \to X_s\mu^+\mu^-$ in several $q^2$ bins; (v) the measurements of differential branching ratio and angular observables of $B_s \to \phi \mu^+\mu^-$ in several $q^2$ bins; (vi)$B_s-\bar{B_s}$ mass difference; (vii) $B^+\to K^+ \nu \bar{\nu}$ branching ratio; (viii) $B^{0,+} \to K^* \mu^+ \mu^-$: CP-averaged angular observables $S_{i = 3,4,5,7,8,9}$, longitudinal polarization fraction of the $K^{0*}$ meson $F_L$, and forward-backward asymmetry of the dimuon system $A_{FB}$,  binned differential branching ratio $dBR/dq^2$; (ix) $B_s^0\to \phi \mu^+ \mu^-$: CP-averaged angular observables $S_{i=3,4,7}$, time-averaged longitudinal polarization $F_L$ fraction, and differential branching ratio $dBR/dq^2$; (x) $B_0 \to K^{*0} e^+e^-$: CP-averaged angular observables $P^\prime_{4,5}$, binned longitudinal polarization fraction $F_L$ and binned differential branching ratio $dBR/dq^2$   for FIT$_1$. 
The FIT$_2$ include all the above-mentioned observables and also  (xi) $B^{0,+} \to K^* \mu^+ \mu^-$:  CP-asymmetries $A_{i=3,4,5,6s,7,8,9}$, binned $A_{CP}$.

All the observables and log-likelihood function are computed with \texttt{flavio}. The best-fit points (BMPs) are obtained by  means of \texttt{Iminuit} package \cite{iminuit} that utilizes the MINOS algorithm \cite{James:1975dr}.

We consider two BMPs originating from the minimization of the log-likelihood function.
The first one (FIT$_1$) corresponds to a scenario with real NP parameters with zero phases:
 \begin{align}
    \alpha_{13} = (2.0\pm 4)\cdot 10^{-3},\qquad \alpha_{23} =-0.207\pm 0.022,\qquad \beta_{13} =0.61\pm 0.10,\nonumber\\ \beta_{23} = 0 \pm 0.5,\qquad M_{Z^\prime}/g_E = 16.1 \pm 0.6 \mathrm{TeV},\nonumber\\
    \phi_{13}=\phi_{23}=\chi_{13}=\chi_{23} = 0,
    \label{eq:FIT1_pars}
 \end{align}
 while for the second one (FIT$_2$), we only nullify the lepton NP phases $\chi_{13}$ and $\chi_{23}$ and allow the quark NP phases $\phi_{13}$ and $\phi_{23}$ to float:  
\begin{align}
	\alpha_{13} = (8\pm 2)\cdot 10^{-3},\qquad \alpha_{23} =0.34\pm 0.08,\qquad \beta_{13} =0.76\pm 0.17,\nonumber\\ \beta_{23} = 0.0 \pm 0.3,\qquad M_{Z^\prime}/g_E = 18.4 \pm 1.7 \mathrm{TeV}, \qquad \phi_{13} = \text{unconstrained},\nonumber\\
    \phi_{23} = 2.49 \pm 0.24,\qquad \chi_{13}=\chi_{23} = 0.
    \label{eq:FIT2_pars}.
 \end{align}
 In the latter case, the fitted value of the mixing angle $\alpha_{13} \sim 0$ is small, so the effect of $\phi_{13}$ is negligible, thus, leaving the latter unconstrained.

The Table \ref{tab:pred_WC} illustrates the predictions for NP Wilson Coefficients evaluated in the $b$-quark mass scale.  It is worth noting that the arguments of all WC corresponding to FIT$_2$ are similar. 
This is due to the $g_L^{bs} \propto \sin \alpha_{23} e^{-i \phi_{23}}$ factor, which accounts for the phases of the NP contributions.  
\begin{table}[t]
\centering
\caption{Values of NP Wilson coefficients for benchmark scenarios FIT$_1$ and FIT$_2$ at the $m_b$  scale}
\label{tab:pred_WC}
 \begin{tabular}{|c|c|c|c|c|c|c|c|c|c|c|c|c|}
			\hline
		 & $C_9^{ee}$ & $C_9^{\prime ee}$ 
		 & $C_9^{\mu\mu}$ & $C_9^{\prime \mu\mu}$ 
		 & $C_9^{\tau\tau}$ & $C_9^{\prime \tau\tau}$ \\
		    \hline
		FIT$_1$  & 
        -0.73 &-0.31 &
		-0.87 &-0.37 &
		-0.58 &-0.24 \\
		FIT$_2$ & 
       -0.62-0.50$i$ & -0.08-0.07$ i$  
		&-0.81-0.65$ i$ &-0.1-0.1$ i$  &
       -0.60-0.48$ i$ &-0.08-0.06$ i$ \\
		\hline\hline
		 & $C_{10}^{ee}$ & $C_{10}^{\prime ee}$& $C_{10}^{\mu\mu}$ & $C_{10}^{\prime \mu\mu}$& $C_{10}^{\tau\tau}$ & $C_{10}^{\prime \tau\tau}$\\
		 \hline
		FIT$_1$ &  -0.16 &-0.06&-0.01 &0.005 & -0.31&-0.12\\
		FIT$_2$ &-0.21-0.16$i$ &-0.026-0.020$i$ &-0.01-0.01$i$&0.001+0.001$i$ & -0.23-0.18$i$ &-0.03-0.03$i$\\
		\hline \hline
		& $C_L^{(\prime)\nu_e \nu_e}$& $C_L^{(\prime)\nu_e \nu_\mu}$ & $C_L^{(\prime)\nu_e \nu_\tau}$& $C_R^{(\prime)\nu_e \nu_e}$& $C_R^{(\prime)\nu_e \nu_\mu}$ & $C_R^{(\prime)\nu_e \nu_\tau}$\\
		\hline
		FIT$_1$ & -0.44 &0.024&-0.02&-0.19&0.01&-0.008\\
		FIT$_2$ & -0.40-0.32$i$&0.07+0.05$i$&-0.05-0.04$i$&-0.05-0.04$i$&0.01+0.01$i$&-0.007-0.005$i$\\
		\hline\hline
		& $C_L^{(\prime)\nu_\mu \nu_\mu}$& $C_L^{(\prime)\nu_\mu \nu_\tau}$ & $C_L^{(\prime)\nu_\tau \nu_\tau}$& $C_R^{(\prime)\nu_\mu \nu_\mu}$& $C_R^{(\prime)\nu_\mu \nu_\tau}$ & $C_R^{(\prime)\nu_\tau \nu_\tau}$\\
		\hline
		FIT$_1$ &-0.17&-0.21&-0.28&-0.07&-0.09&-0.12\\
		FIT$_2$ &-0.15-0.12$i$&-0.18-0.16$i$&-0.28-0.22$i$&-0.021-0.016$i$&-0.025-0.021$i$&-0.038-0.030$i$\\
		\hline
		\end{tabular}
\end{table}	

\begin{table}[h]
\centering
\caption{Predictions for several $b\to s$ observables. Here $\Delta = \frac{|A_{pred} - A_{exp}|}{\sqrt{\Delta A_{pred}^2 + \Delta A_{exp}^2}}$. Note, that the world average result for $\mathcal{B}(B^+ \to K^+\nu \bar{\nu})$ is interpreted as an experimental measurement when calculating the corresponding $\Delta$. }
\label{tab:pred}
\scriptsize{
 \begin{tabular}{|c|c|c|c|c|c|c|}
			\hline
		Obs & SM & Exp & FIT 1 & $\Delta_1$& FIT 2 &$\Delta_2$\\
		    \hline
		$R_K(B^+)^{[1.1, 6.0]}$ & $1\pm 0.01$  \cite{Hiller:2003js},\cite{Bordone:2016gaq},\cite{Isidori:2020acz} & $0.949^{+0.042}_{-0.041}\pm 0.022$ \cite{LHCb:2022zom}& $0.896\pm 0.015$ & 0.81& $0.898\pm 0.020$ &0.76 \\
		\hline
		$R_K^*(B^0)^{[1.1, 6.0]}$ &  $1\pm 0.01$ \cite{Hiller:2003js},\cite{Bordone:2016gaq} & $1.027^{+0.072}_{-0.068}  \pm  0.027$ \cite{LHCb:2022zom}& $0.955\pm 0.012 $ &0.73 &$0.923\pm 0.015$ & 1.05\\
		\hline
		$P_5^{\prime[4,6]}$ & $-0.757 \pm 0.077$ \cite{Descotes-Genon:2013vna}&$-0.439 \pm 0.111  \pm 0.036$\cite{LHCb:2020lmf}& $-0.53 \pm 0.09 $ &0.53&  $-0.55 \pm 0.05$&0.76\\
		\hline
		$\Delta M_{B_s}, \mathrm{ps^{-1}}$ & $18.77 \pm 0.76$ \cite{Lenz:2019lvd}& $17.765 \pm 0.004$ \cite{HFLAV:2022pwe}&$17.94 \pm 2.76$ &0.07& $18.48 \pm 2.10$ &0.35\\
		\hline
		$\mathcal{B}(B_s\to \mu\mu)\cdot 10^{-9}$  & $3.68\pm 0.14 $ \cite{Beneke:2019slt}& $3.09^{+0.46+0.15}_{-0.43-0.11} $\cite{LHCb:2021vsc}&$3.69\pm 0.21$ & 1.02& $3.68\pm 0.16 $& 1.04\\
		\hline \hline
		$\mathcal{B}(B^+ \to K^+\nu \bar{\nu})\times 10^{-6}$ &  $4.6\pm 0.5$ \cite{Belle-II:2018jsg}& $11 \pm 4$\cite{Dattola:2021cmw},$<19$\cite{Belle:2017oht}  & $5.03 \pm 0.67$&1.46&$4.88 \pm 0.64$&1.51\\
         \hline
         $\mathcal{B}(B^0 \to K^0\nu \bar{\nu})\times 10^{-6}$  & $4.1\pm 0.5$ \cite{david_straub_2021_5543714}& $<26$  \cite{Belle:2017oht}&$4.65 \pm 0.92$ &&$4.51\pm 0.70$&\\
         \hline
         $\mathcal{B}(B^0 \to K^{0*}\nu \bar{\nu})\times 10^{-6}$  &  $9.6\pm 0.9$ \cite{Belle-II:2018jsg}& $<18$  \cite{Belle:2017oht}&$ 10.20 \pm 0.96$ &&$10.30 \pm 1.02$&\\
         \hline
         $\mathcal{B}(B^+ \to K^{+*}\nu \bar{\nu})\times 10^{-6}$ &  $9.6\pm 0.9$ \cite{Belle-II:2018jsg}& $<61$  \cite{Belle:2017oht} &$11.00 \pm 0.90$ &&$ 11.10 \pm 1.20$&\\
         \hline
         $F_L^{B^0 \to K^{*}\nu \bar{\nu}}$ &  $0.47 \pm 0.03$ \cite{Buras:2014fpa}&- &$0.465\pm0.04$&  & $0.469 \pm 0.05$  & \\
         \hline
         $R_K^{\nu \bar{\nu}}$ &  1 & $2.4\pm 0.9 $ &$1.14\pm 0.028$ & & $1.11\pm 0.024$ & \\
         \hline
         $R_{K^*}^{\nu \bar{\nu}}$ &  1& $<1.9$ & $1.07\pm 0.024$ && $1.09\pm 0.022$ &  \\
         \hline 
		\end{tabular}
		}
\end{table}	

The Table ~\ref{tab:pred} lists the model's predictions for various observables considered in this paper. In appendix \ref{sec:CPaveraged} we also present the comparison between the experiment, the SM, and our model for CP-averaged angular coefficients (Tabs.~\ref{tab:pred_S3456},\ref{tab:pred_S789}).  
Note that we consider two bins of $q^2$: $q^2 \in [1.1, 6]$ (central-$q^2$) and $q^2 \in [15, 19]$ (high-$q^2$).  To avoid the charmonium resonances, we refrain from making any prediction in the $ [6, 15]$ GeV$^2$ range.
One can see that  FIT$_1$ and FIT$_2$ are compatible both with the SM and the current experimental measurement, slightly relaxing some of the SM discrepancies, e.g. for $P'_5$ ($S_5$) in lower $q^2$ bins.

The predictions of $A_{CP}(K^{(*)})$ in $B\to K^{(*)}\mu^+\mu^-$ together with triple-product asymmetries $A_{7,8,9}$ for all fits are given in Table \ref{tab:pred_A7_ACP}. 
The  $A_{3,4,5,6s}$  CP asymmetries to be less or almost  a percent in the central- and high-$q^2$ bin for both fits, respectively,  and hence making their observation a difficult attempt, therefore, we did not present it.

\begin{table}[h]
\centering
\caption{Prediction of certain angular CP asymmetries   in $B^0\to K^* \mu^+\mu^-$ and $B^+\to K^+ \mu^+\mu^-$ in the central- and high-$q^2$ region.}
\label{tab:pred_A7_ACP}
 \begin{tabular}{|c|c|c|c|c|c|}
			\hline
		 & $A_7^{[1.1,6]}$(\%)& $A_8^{[1.1,6]}$(\%)& $A_9^{[1.1,6]}$(\%) &$A_{CP}^{[1.1,6]}(K^*)$(\%)&$A_{CP}^{[1.1,6]}(K)$(\%)\\
		    \hline
		    EXP \cite{LHCb:2015svh}&$-4.5^{+5.0}_{-5.0}\pm 0.6$&$-4.7^{+5.8}_{-5.7}\pm 0.8$&$-3.3^{+4.0}_{-4.2}\pm 0.4$& $-9.4\pm 4.7$ \cite{LHCb:2014mit}& $0.4\pm 2.8$\cite{LHCb:2014mit}\\
		\hline
		FIT$_1$ &$0.28 \pm 0.12$&$-0.19 \pm 0.34$&$-0.01 \pm 0.07$&$0.06 \pm 0.04$&$0.06 \pm 0.07$\\
		\hline
		FIT$_2$ &$0.43 \pm 0.06$& $\mathbf{-2.59 \pm 0.27}$&$-0.24 \pm 0.02$&$0.11\pm 0.04$&$-0.29\pm 0.05$ \\
		\hline\hline
	& $A_7^{[15,19]}$(\%)& $A_8^{[15,19]}$(\%)& $A_9^{[15,19]}$(\%) &$A_{CP}^{[15,19]}(K^*)$(\%)&$A_{CP}^{[15,19]}(K)$(\%)\\
		    \hline
		EXP \cite{LHCb:2015svh} &$-4.0^{+4.5}_{-4.4}\pm 0.6$&$2.5^{+4.8}_{-4.7}\pm 0.3$&$6.1^{+4.3}_{-4.4}\pm 0.2$&$-7.4\pm 4.4$ \cite{LHCb:2014mit}& $-0.5\pm 3.0$ \cite{LHCb:2014mit}\\
		\hline
		FIT$_1$ &$0.01 \pm 0.04$&$-0.04 \pm 0.10$&$-0.06 \pm 0.11$&$-0.23 \pm 0.17$&$-0.48 \pm 0.40$\\
		\hline
		FIT$_2$ &  $0.017 \pm 0.03$&$-0.44 \pm 0.13$&$-0.69 \pm 0.19$& $-1.40 \pm 0.28$&$\mathbf{-3.21\pm 0.61}$\\
		\hline
		\end{tabular}
\end{table}	
It is apparent that none of the new physics fits can enhance  $A_{CP}(K^{(*)})$ in the central-$q^2$ bin at the level of a few percent.  However, such an enhancement is feasible in the high-$q^2$ region for FIT$_2$.  Here one should emphasize that although the enhancement in the high-$q^2$ bin is more prominent, the measurement of $A_{CP}$ in the central-$q^2$ region appears to be more attractive as the branching ratio in the central-$q^2$ region is larger as compared to the high-$q^2$ bin. The LHCb,  Belle-II experiment is expected to collect a sample of a few thousand events of $B^0\to K^*\mu^+\mu^-$ \cite{Belle-II:2018jsg}, \cite{Lopes:2005km}, allowing a measurement of the branching ratio and its CP asymmetry at the percent level. 

\begin{figure}[h]
		\begin{tabular}{ccc}
        \includegraphics[width=0.33\textwidth]{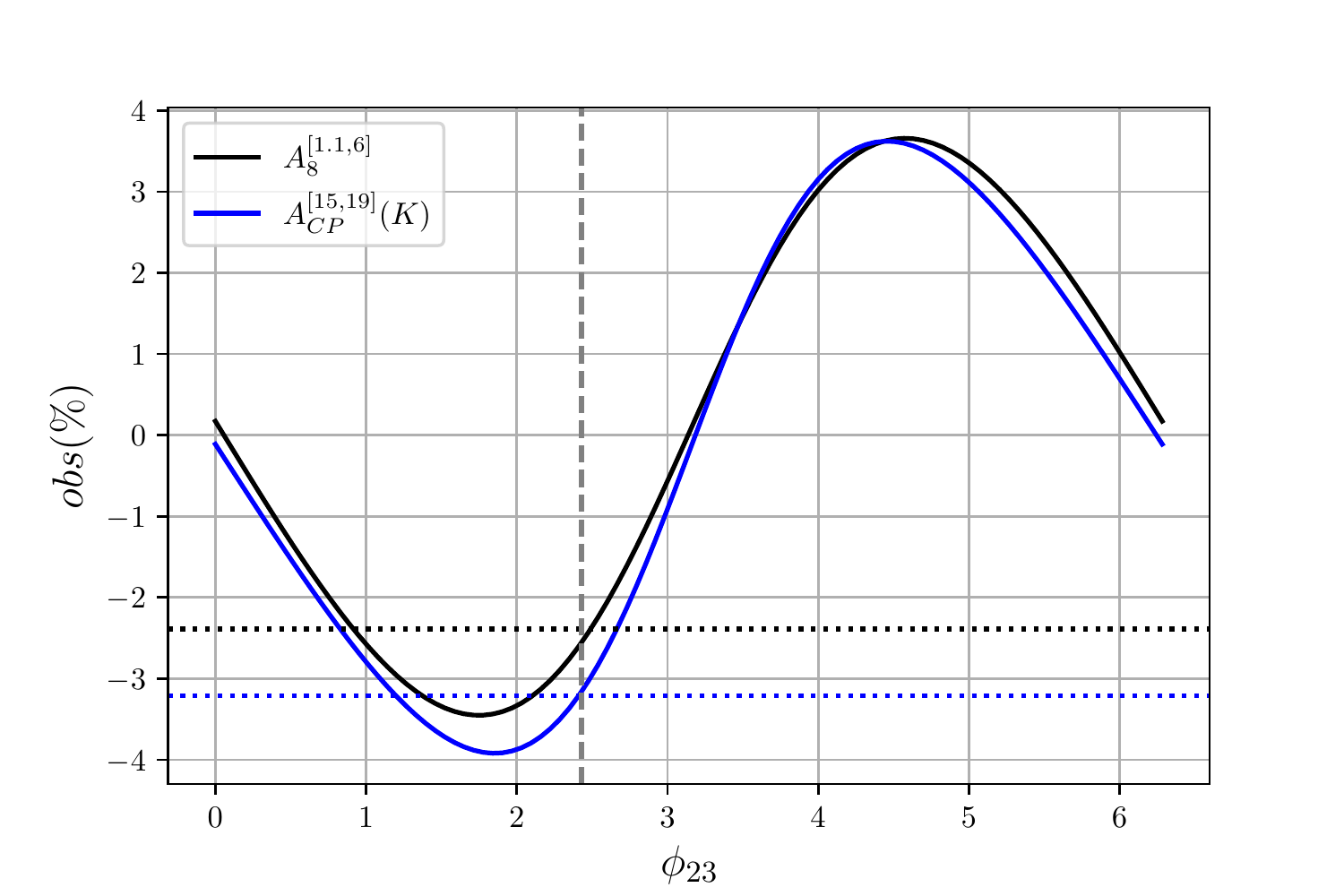}&
        \includegraphics[width=0.33\textwidth]{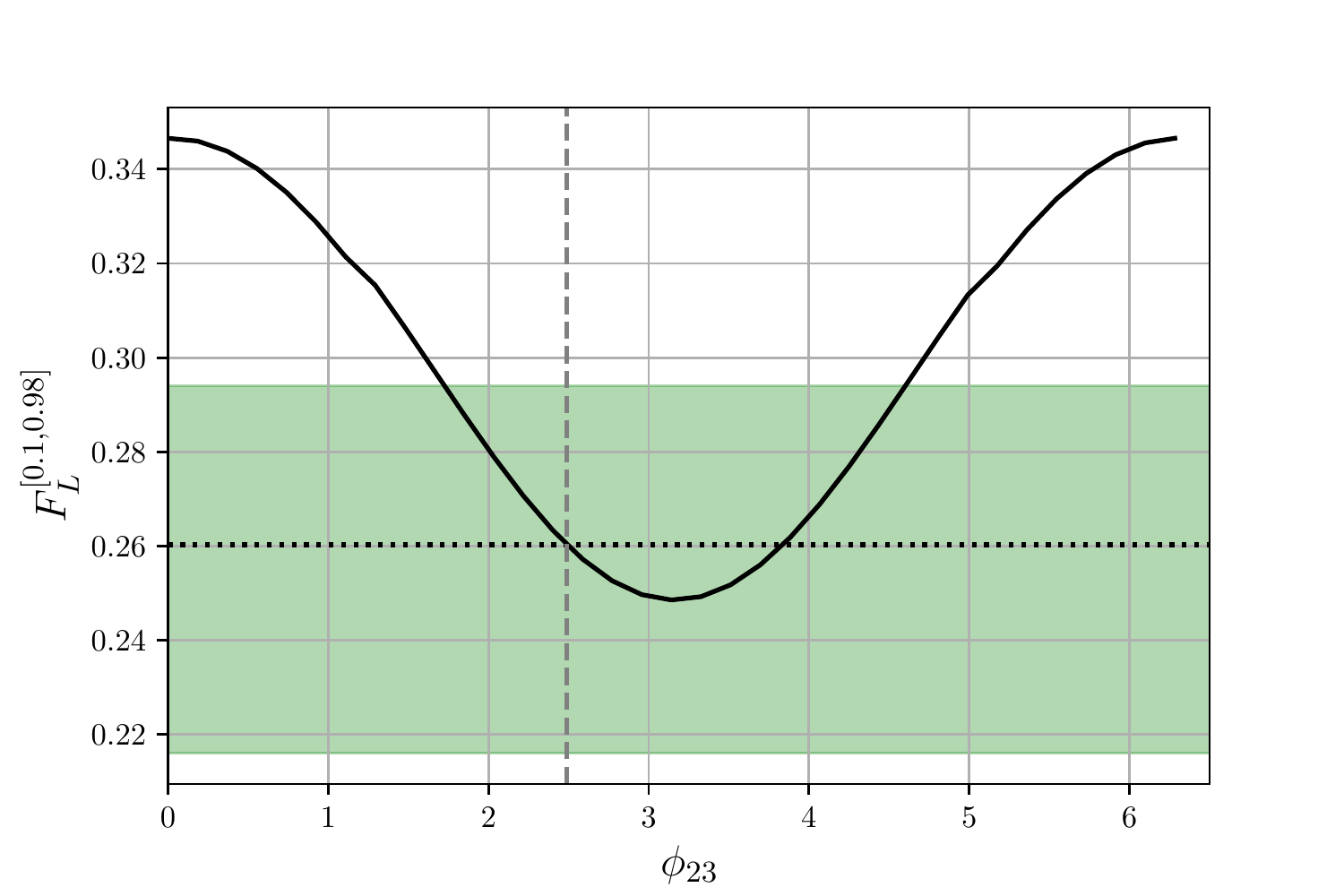}&
        \includegraphics[width=0.33\textwidth]{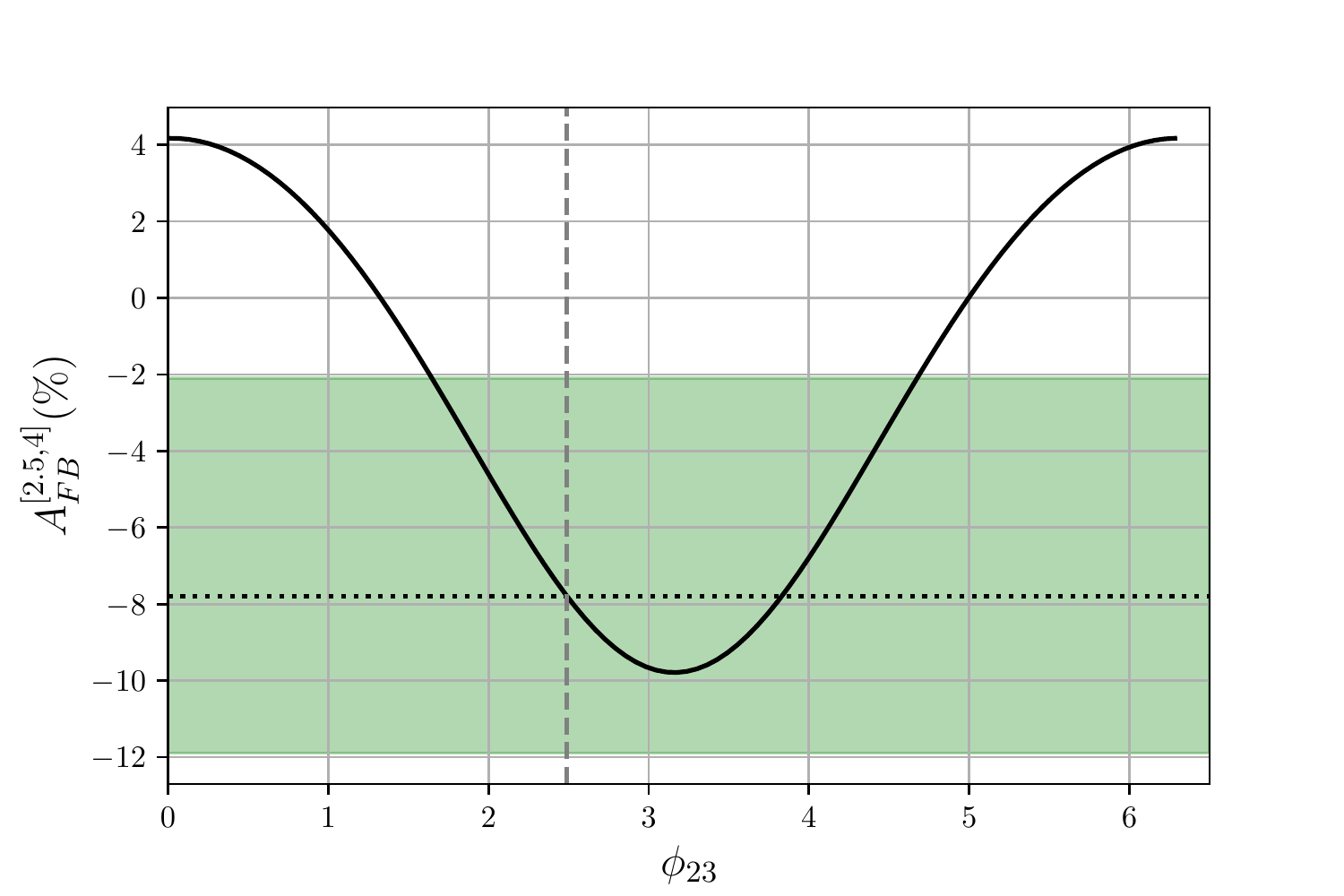}
    	\end{tabular}
	\caption{The new weak phase dependence of the  $A_8^{[1.1,6]}$, $A_{CP}^{[15,19]}(K)$, $F_L^{[0.1,0.98]}$, $A_{FB}^{[2.5,4]}$ observables. Here green band is $1\sigma$ experimental limit \cite{LHCb:2020lmf}. Dotted line is central value of model prediction for FIT$_2$. }
\label{fig:phase_var}
\end{figure}

In the Ref.~\cite{Biswas:2020uaq} authors considered $O_{9,10}$ operators with both real and complex WCs. They have pointed out imaginary contributions arising in  the CP-averaged and CP-asymmetric observables.
For example, they mention that $A_8^{[1.1,6]}$  favor negative values of $Im(\Delta C_9)$, though positive values are also possible; the $B\to K^{(*)} ll$ observables $F_L^{[0.1,0.98]}$, $A_{FB}^{[2.5,4]}$ can be also potentially enhanced due to a non-zero contribution from imaginary $\Delta C_9$. 

From  Tab.~\ref{tab:pred_WC} one can see that, indeed, our FIT$_2$ favors negative imaginary parts for $C^{(')}_9$. For convenience, we provide Fig.\ref{fig:phase_var}, where we depict the $\phi_{23}$ dependence of the above-mentioned $A_8^{[1.1,6]}$,  $F_L^{[0.1,0.98]}$, $A_{FB}^{[2.5,4]}$ together with $A_{CP}^{[15,19]}(K)$ for $B^+\to K^+ \mu^+\mu^-$ in the case when all other parameters are fixed according to FIT$_2$. From this figures it can be observed that the $A_8^{[1.1,6]}$, $A_{CP}^{[15,19]}(K)$ asymmetries can attain values of 3.5\%. Differences from zero of any value of it would be an unambiguous indication of the existence physics beyond the SM.  One can also see that in our scenarios $F_L^{[0.1,0.98]}$, $A_{FB}^{[2.5,4]}$ lie well within the 1$\sigma$ experimental limits \cite{LHCb:2020lmf} indicated as green bands. The dotted lines on these figures correspond to the model prediction for FIT$_2$. 
Other $A_i$ asymmetries are $\sim 1\%$ or less, and we do not show them.

\begin{figure}[h]
\centering
  \includegraphics[width=0.5\textwidth]{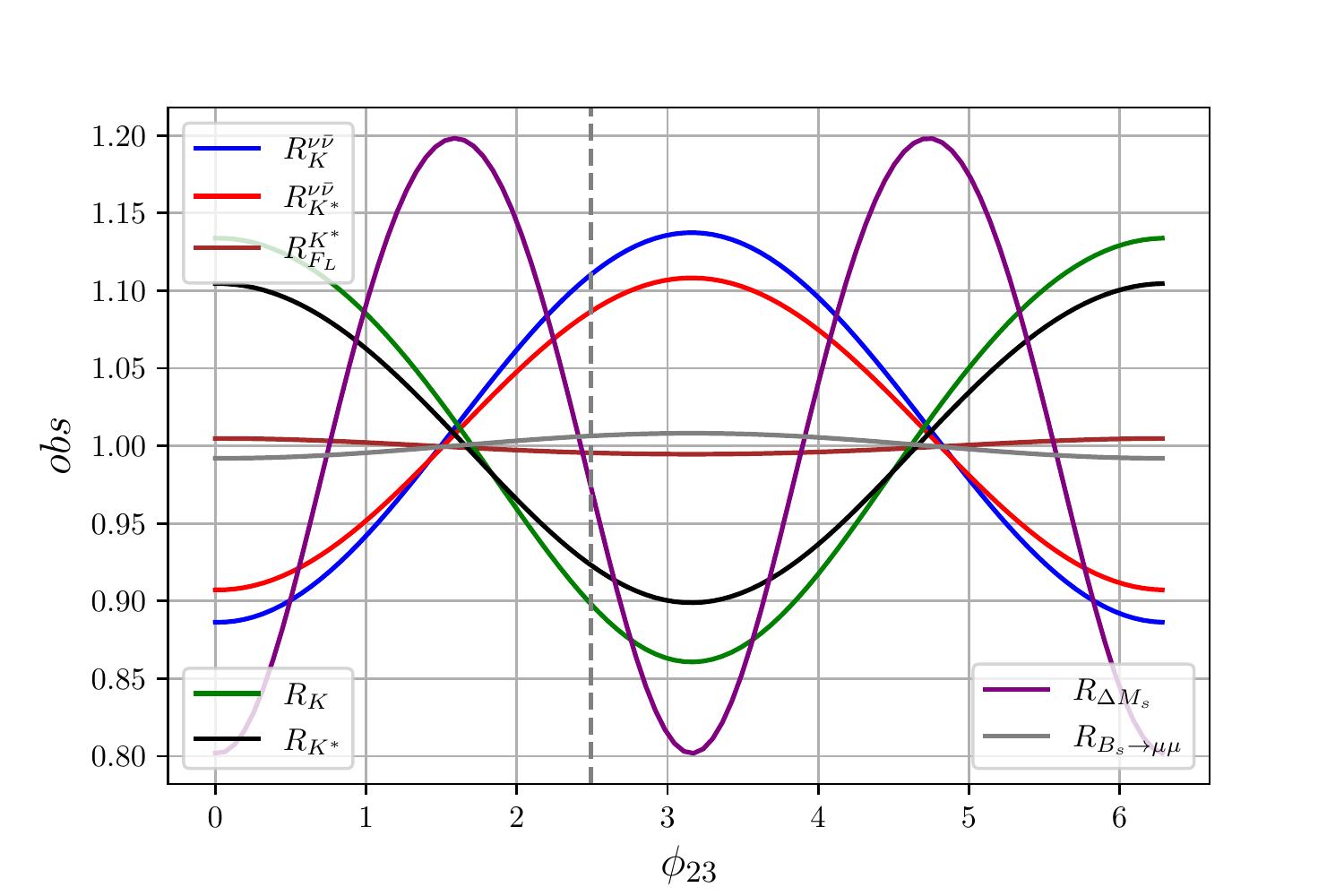}
	\caption{ Phase dependence for $b\to s$ obervables}
\label{fig:phase_var_RK}
\end{figure}

It is also interesting to demonstrate how other key observables can be enhanced/suppressed w.r.t. the SM with $\phi_{23}$. In Fig.~\ref{fig:phase_var_RK} we plot $R_{K^{(*)},F_L}^{\nu\bar{\nu}}$ defined in Eq.~\eqref{eq:Rnunu}, together with  analogous ratios for $\Delta M_s$ and $\mathcal{B}(B_s \to \mu\mu)$. In addition, the dependence on $\phi_{23}$ of lepton-flavour violation ratios $R_K$ and $R_{K^*}$ in charged-lepton channel is shown. One can see that indeed the sign of $(R^{\nu\bar{\nu}}_{F_L} - 1)$ is opposite to that of $(R^{\nu\bar{\nu}}_{K^*} - 1)$ (see the discussion at the end of Sec.~\ref{sec:BKnunu}).

The Wilson coefficients corresponding to the left-quark chiral currents by definition ~\eqref{eq:Heff} are $C_{9,10}$ for $b\to sll$ and $C_L^{(\prime)}$ for $b\to s\nu\bar{\nu} $ transitions. Similarly, WCs corresponding to the right-quark chiral currents ~\eqref{eq:Heff}, which exist purely in beyond the SM scenarios,  are  $C_{9,10}^\prime$ and  $C_R^{(\prime)}$ for $b\to s ll$ and  $b\to s\nu\bar{\nu} $, respectively.
Since the factors $g_{L(R)}^{bs}$ enter all the considered Wilson coefficients, there exist certain relations between WC, e.g., for any fixed $\nu,\nu',l,l'$ we have
\begin{align}
	g^{bs}_{R}/g^{bs}_{L} = C_R^{\nu\nu'}/C_L^{\nu\nu'} = C_R^{'\nu\nu}/C_L^{\nu\nu'} = C_9^{'ll}/C_9^{ll'} = C_{10}^{'ll}/C_{10}^{ll'} = \BbsR,
\end{align}
where the factor 
\begin{align}
	\BbsR =  \frac{m_b m_s m_d^2}{m_d^2 (m_b^2 \sin^2 \alpha_{23} + m_s^2 \cos^2 \alpha_{23})\cos^2 \alpha_{13}+m_b^2m_s^2\sin^2 \alpha_{13}}
\label{eq:bsR_factor}
\end{align}
given in Ref.~\cite{Bednyakov:2021fof} is modified to account for both NP quark angles.
Clearly, such kind of relations lead to certain imprint in the predictions for observables.

We discuss interdependencies between different observables for the two cases: without and with new CP-phases.  To carry out the analysis, first of all we calculate the best fit points as discussed earlier in Sec.\ref{sec:fit}.  After that we randomly generate a list of model parameters for $1(3)\sigma$ variations near our BMPs, and calculate model predictions for 
various observables. 

We present the results of the study in a form of two-dimensional scatter plots only for the observables that exhibit largest deviations in either bin.

\begin{figure}[h]
		\begin{tabular}{ccc}
		\includegraphics[width=0.33\textwidth]{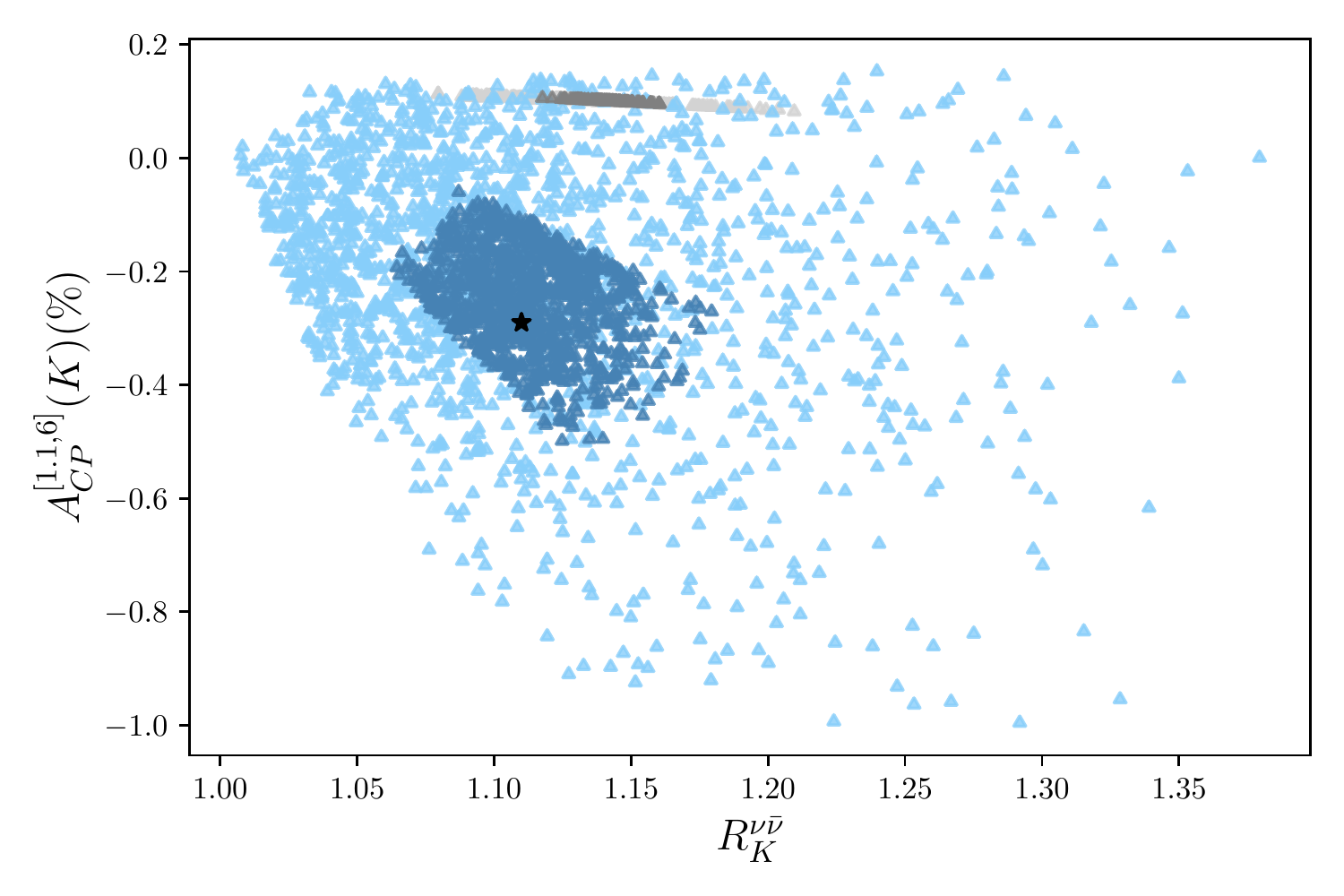}&
        \includegraphics[width=0.33\textwidth]{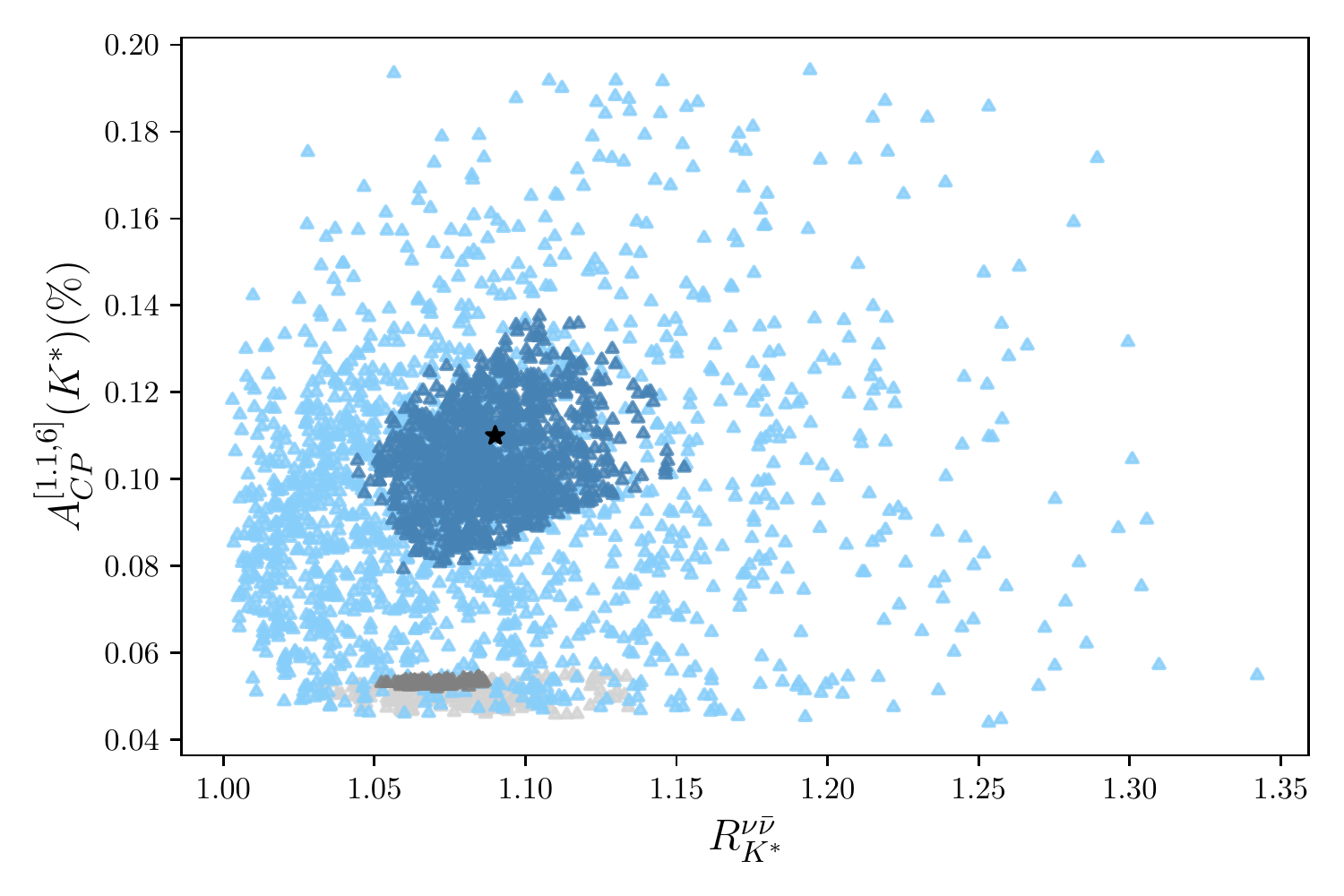}&
        \includegraphics[width=0.33\textwidth]{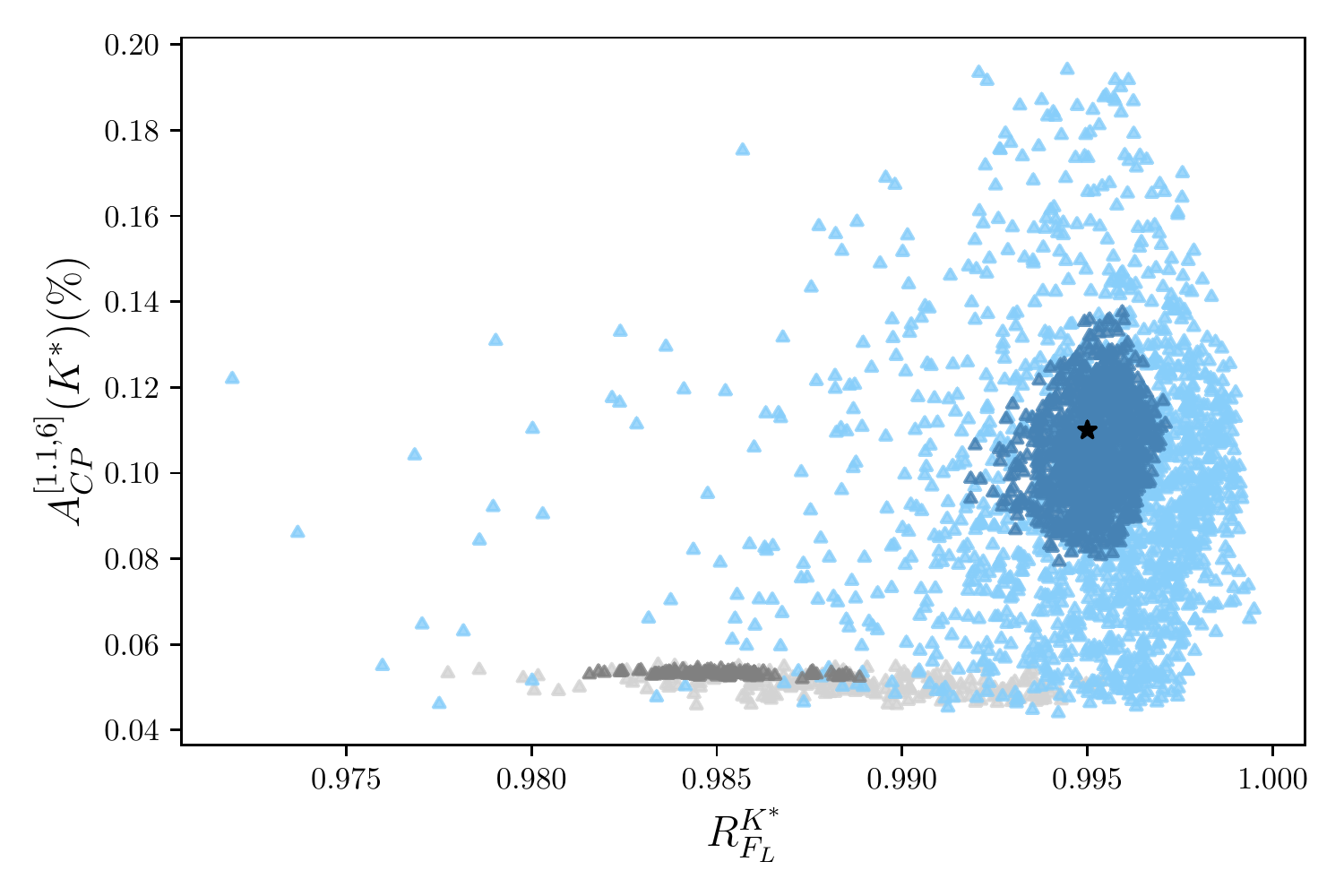}\\
        \includegraphics[width=0.33\textwidth]{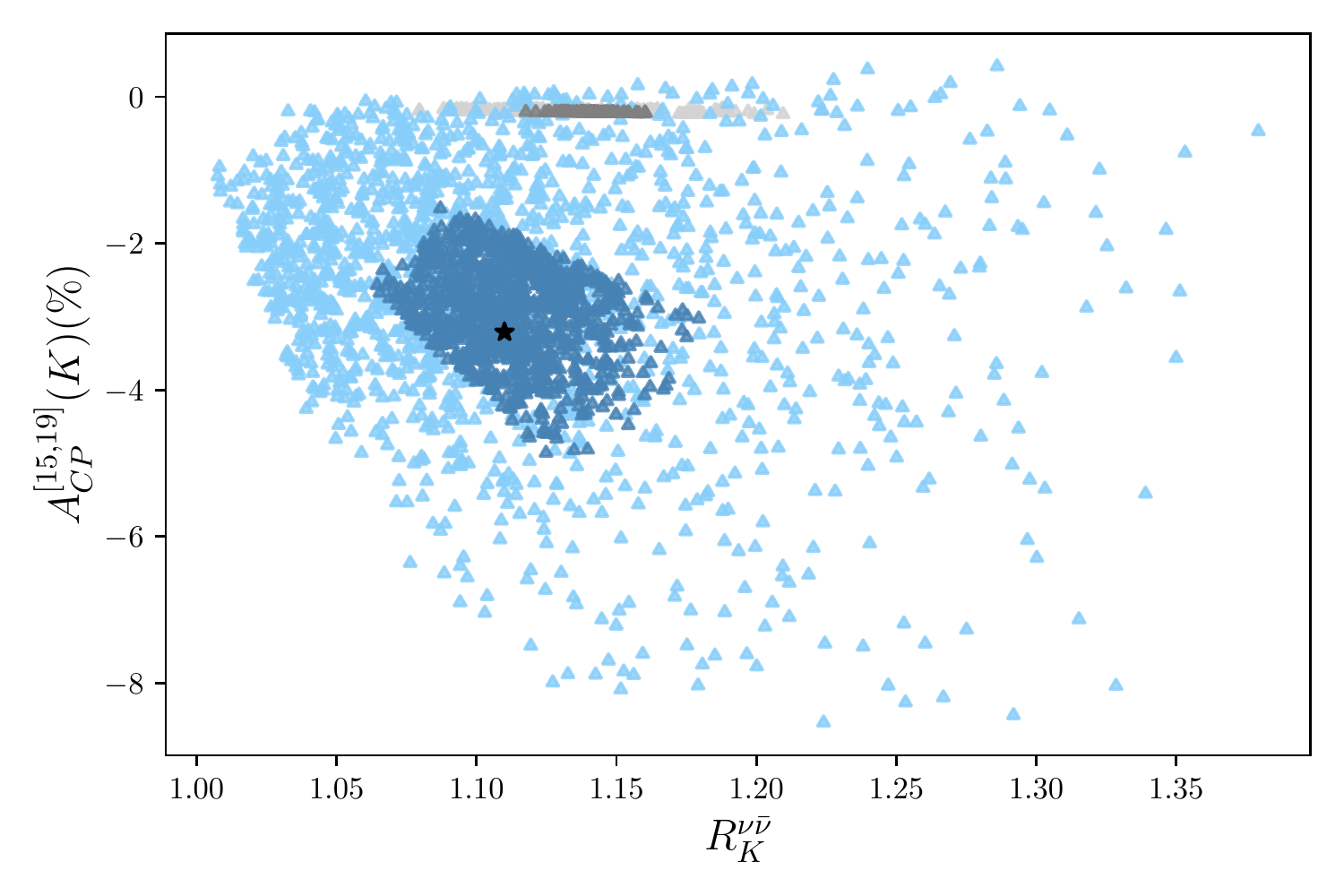}&
        \includegraphics[width=0.33\textwidth]{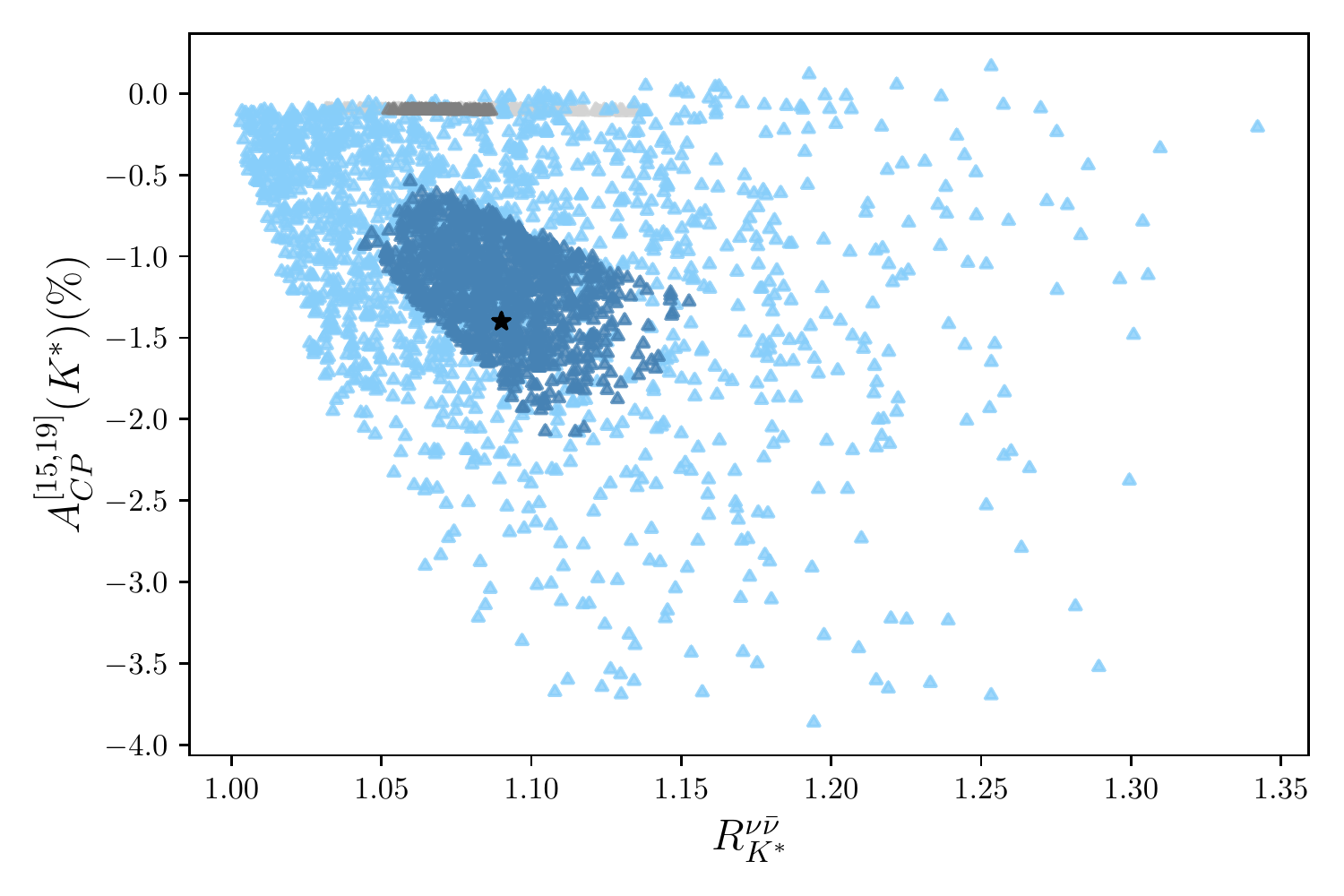}&
        \includegraphics[width=0.33\textwidth]{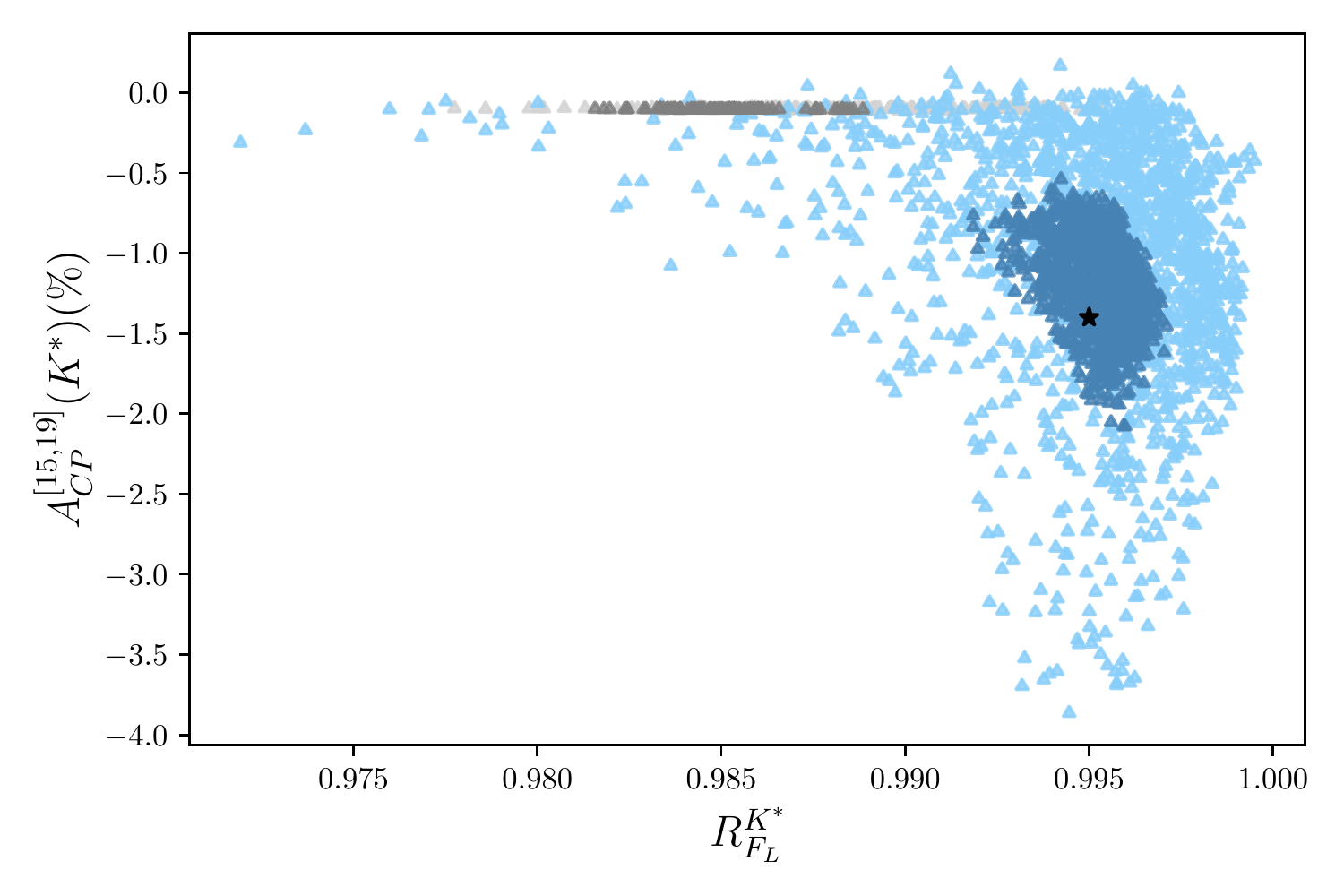}
	\end{tabular}
	\caption{The first row defines dependencies between $A_{CP}(K^{(*)})$ and $R_K^{\nu\bar{\nu}}$, $R_{K^*}^{\nu\bar{\nu}}$, $R_{F_L}^{\nu\bar{\nu}}$  in the central-$q^2$ region, and the second row in the high-$q^2$ region. 
	Gray and lightgray are $1,3\sigma$ variation around central values of model parameters for FIT$_1$; blue and skyblue for FIT$_2$.  The black star is our BMP for FIT$_2$.}
\label{fig:RKnu_ACP}
\end{figure}

In Fig.\ref{fig:RKnu_ACP} we show $A_{CP}(K^{(*)})$ for $B \to K^{(*)}ll$ together with $b\to s\nu\bar{\nu}$ observables discussed in Sec.\ref{sec:BKnunu}. . 
We see that given 3$\sigma$ model parameter variation, $R_K^{\nu\bar{\nu}} \in (1.0-1.35)$, $R_{K^*}^{\nu\bar{\nu}} \in (1.0-1.35)$, and $R_{F_L}^{\nu\bar{\nu}} \in (0.975-1.000)$.  For central-$q^2$ region we have rather small $A_{CP}(K) \sim 0.2-1.0\%$ and even smaller $A_{CP}(K^*) \sim  0\%$.  On the contrary, for high $q^2$-region $A_{CP}(K) \sim 1-8\%$, while $A_{CP}(K^*) \sim  0.2-4\%$.

\begin{figure}[h]
		\begin{tabular}{ccc}
		\includegraphics[width=0.33\textwidth]{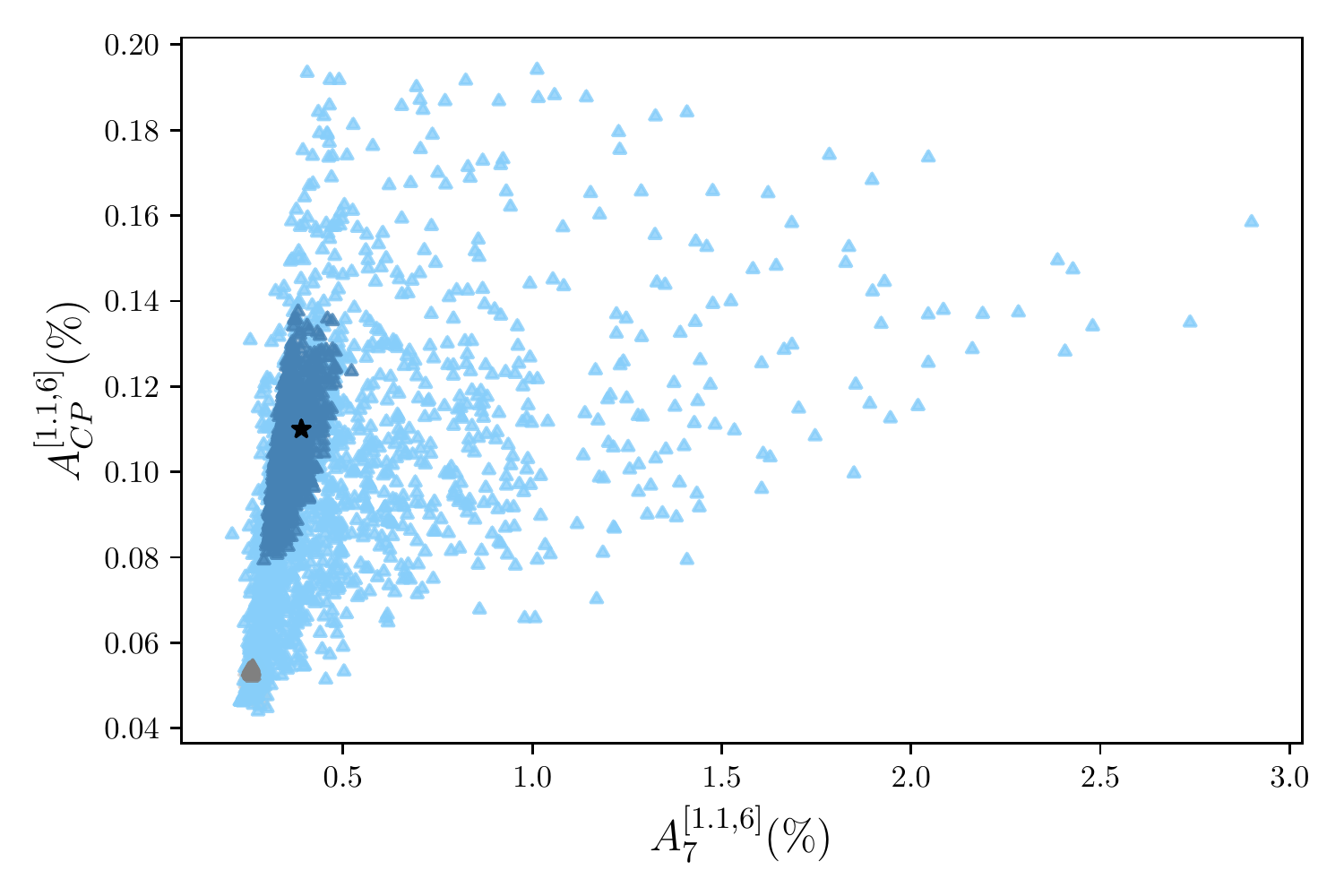}&
        \includegraphics[width=0.33\textwidth]{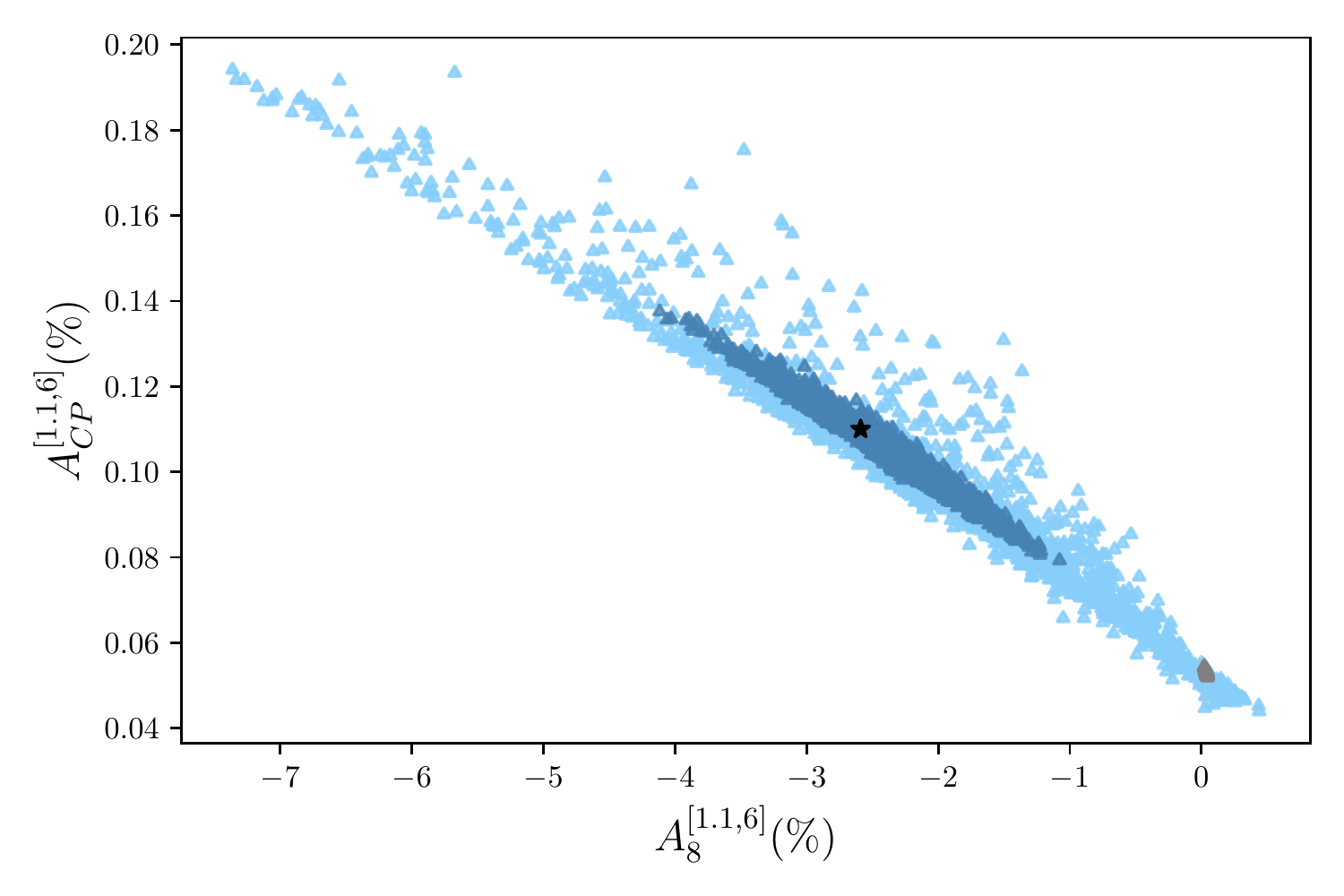}&
        \includegraphics[width=0.33\textwidth]{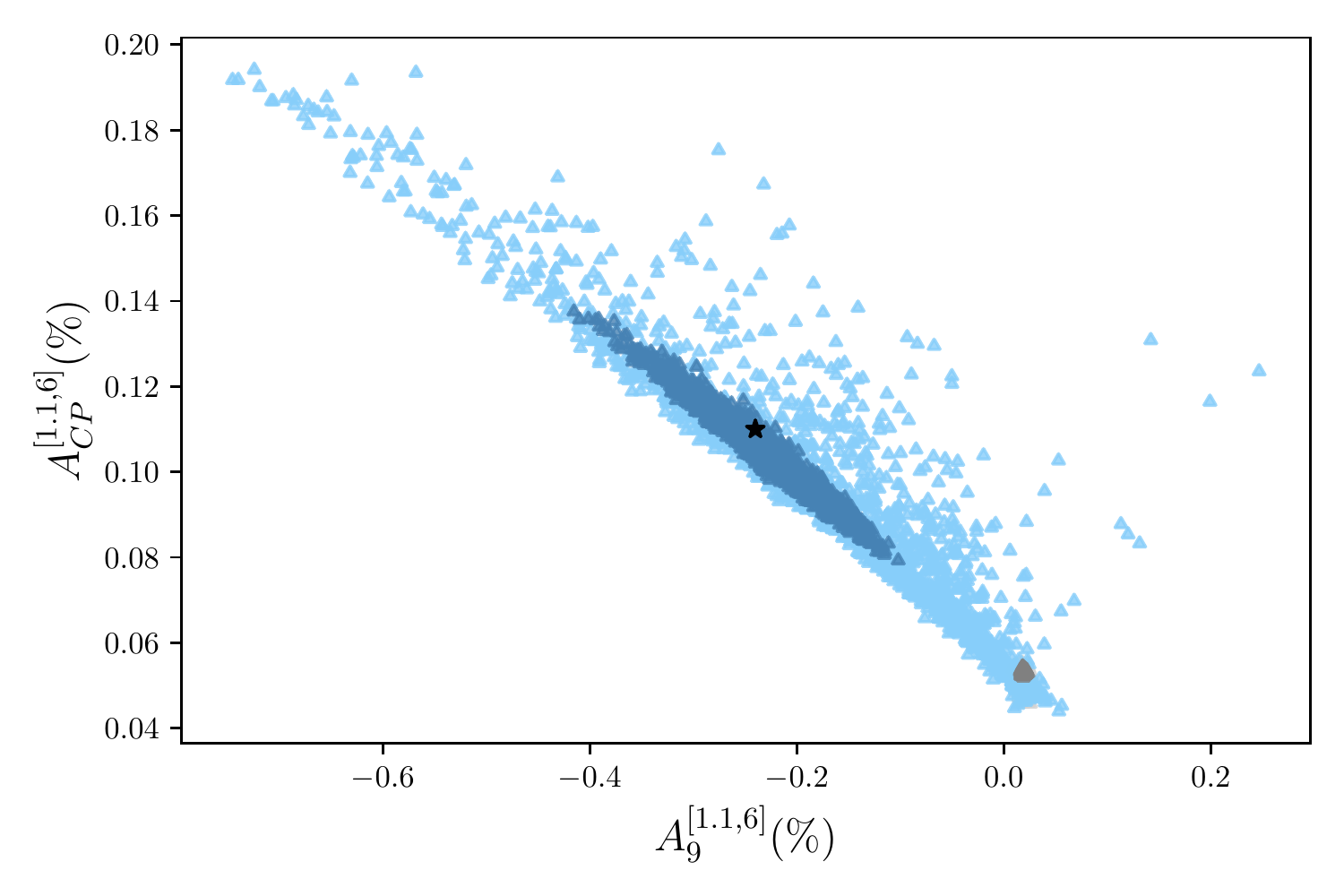}\\
        \includegraphics[width=0.33\textwidth]{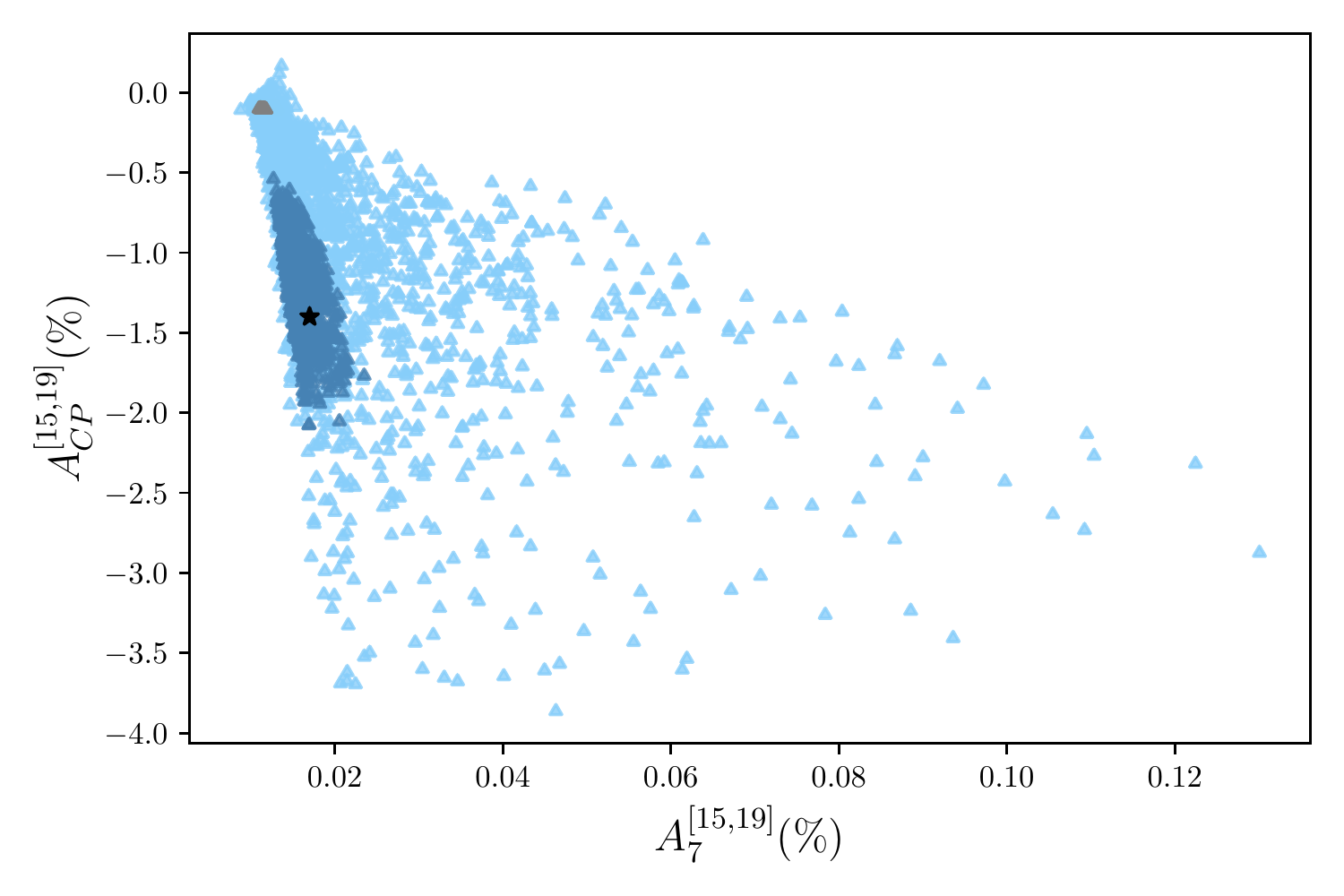}&
        \includegraphics[width=0.33\textwidth]{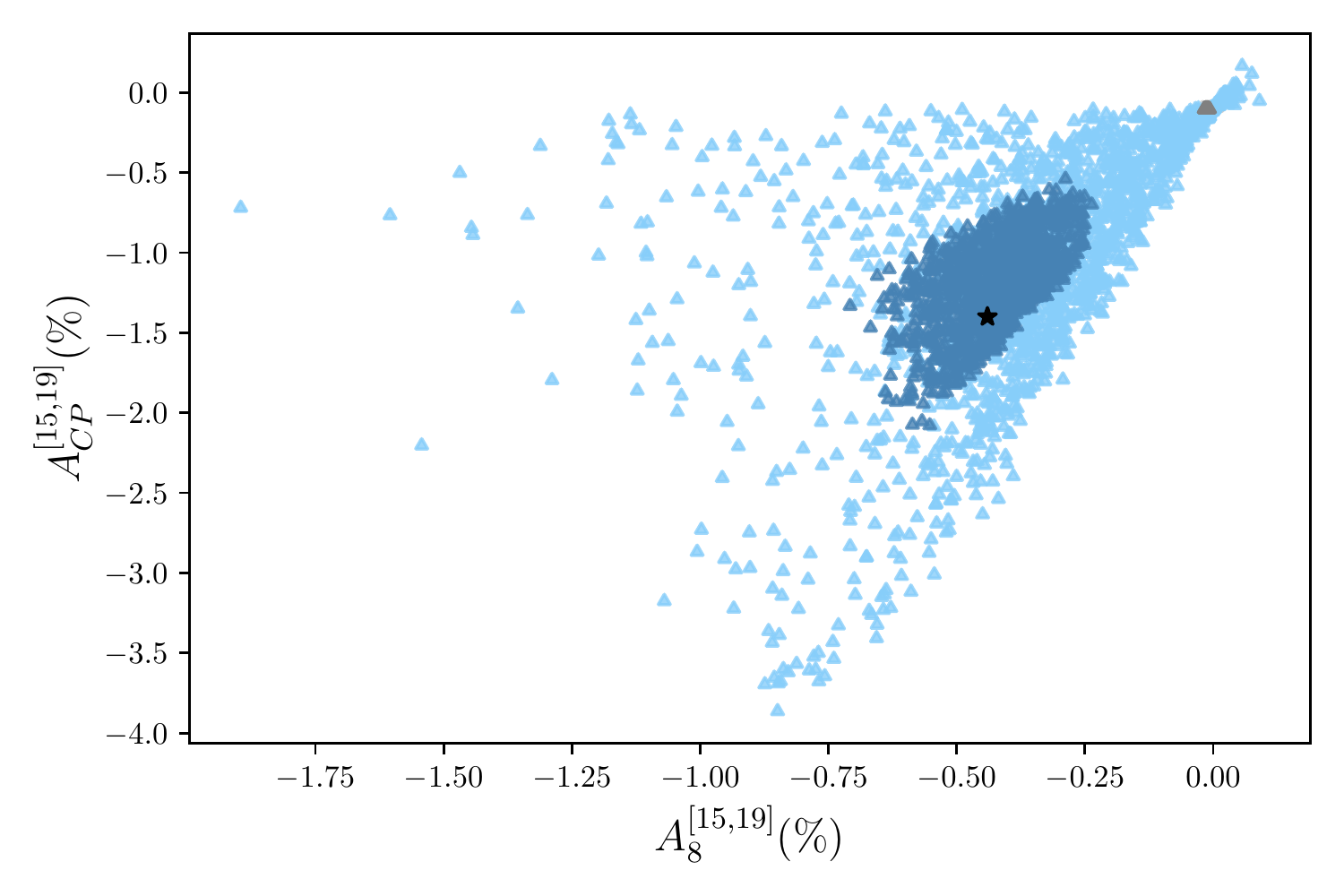}&
        \includegraphics[width=0.33\textwidth]{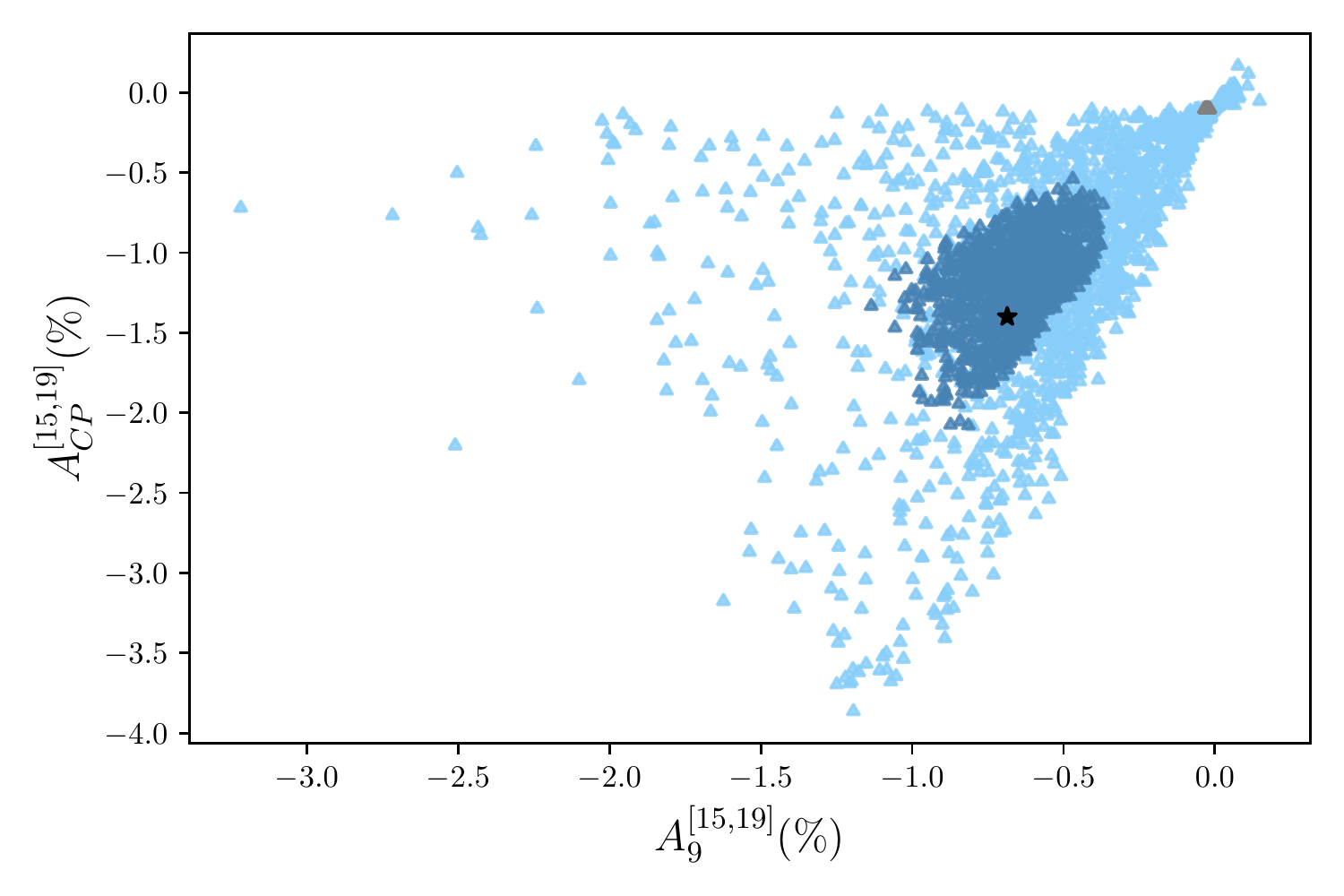}
	\end{tabular}
	\caption{The first (second) row panel portrays dependencies between $A_{CP}$ in  $B^{0}\to K^{*} \mu^+\mu^-$ and $A_7$, $A_8$, $A_9$ in the central-$q^2$ and high-$q^2$ region, respectively. 
	Gray and lightgray are $1,3\sigma$ variation around central values of model parameters for FIT$_1$; blue and skyblue for FIT$_2$.  The black star is our BMP for FIT$_2$.}
\label{fig:A789_ACP}
\end{figure}

For a finer investigation of possible physics contributions to $b\to sll$, we study the correlations between $A_{CP}$ and triple product $A_{7,8,9}^{[1.1,6],[15,19]}$ (see Fig.\ref{fig:A789_ACP}) as for all other angular observables, enhancements are too small to be observed in near future. 

For FIT$_2$ even if $A_{CP}^{[1.1,6]}$ is extremely small, $A_7^{[1.1,6]}$ can still have a large value. For e.g., for $A_{CP}^{[1.1,6]} \sim 0\%$, $A_7^{[1.1,6]} \sim 0.2-2.5\%$. 
The interdependencies between $A_{CP}^{[1.1,6]}$ and $A_8^{[1.1,6]}$ are illustrated in the second  panel of Fig.\ref{fig:A789_ACP}. For FIT$_2$   $A_8^{[1.1,6]}\sim 1-8\%$ for $A_{CP}^{[1.1,6]} \sim 0\%$. For example, for $A_{CP}^{[1.1,6]} \sim 0.19\%$, $A_8^{[1.1,6]}$ can be as large as $-7\%$. Further, there is an anti-correlation between these two observables, i.e., an increase in $A_{CP}^{[1.1,6]}$ would result in decrease in the value of $A_8^{[1.1,6]}$. 
At the end, the correlations between $A_{CP}^{[1.1,6]}$ and $A_9^{[1.1,6]}$ are explained in the third  panel of Fig.\ref{fig:A789_ACP}. There are for FIT$_2$ $A_9^{[1.1,6]}\sim 0-1\%$
for $A_{CP}^{[1.1,6]} \sim 0\%$. 

Now we consider predictions of $A_{7,8,9}$ observables in the high-$q^2$ bin as given in Table  \ref{tab:pred_A7_ACP}. The most distinguishing feature of predictions in the high-$q^2$ region is related to the observable $A_8^{[15,19]}$ which hinted to be a potential observable to phrase the signatures of NP phase. 
FIT$_2$ appears to make a decisive impact as it can enhance $A_{8}^{[15,19]}$ observable up to a level  of $-1.5 \%$ and $A_{9}^{[15,19]}$ up to a level of $-2.5 \%$.   
As for $A_7^{[15,19]}$, FIT$_2$ can not enhance it. 
At the same time we observe the enhancement in $A_{CP}^{[15,19]} \sim 0.1-4\%$ in all cases with high-$q^2$. 

Let us close this section by mentioning the fact that, while our fits \eqref{eq:FIT1_pars} and \eqref{eq:FIT2_pars} for flavour observables constrain only the ratio $M_{Z'}/g_E$, we have checked that, e.g., for $M_{Z'}=3.2$ TeV, one can satisfy the constraints due to negative results of $pp \to X \to ll$ searches \cite{ATLAS:2019erb} and the absence of Landau pole for the $Z'$ gauge coupling $g_E(1~\mathrm{TeV})\leq 0.3$ (see, e.g., Ref.~\cite{Duan:2018akc})

\section{Conclusion \label{sec:res}}
Very interesting deviations from the SM predictions have been found in $b\to s$ transitions. In this article we studied these puzzles in a simplified framework involving a heavy $Z^\prime$ boson. We derived the flavour structure of such model with additional U(1) symmetry, which includes new parameters, such as quark (lepton) mixing angles and a complex phase $\phi_{23}$, which enters $b\to s$ transitions. 

In the phenomenological part of this paper, we first  presented the benchmark points, which are capable of providing a common explanation of all the flavour data. In particular, sizeable CP violation in $B^0\to K^*\mu^+\mu^-$ observables, for example, in $A_8^{[1.1,6]}$, $A_{CP}^{[15,19]}(K)$ and $A_{CP}^{[15,19]}(K^*)$, is predicted.  We also explore the new weak phase dependence of different observables. 

Then we analyzed relations  between $b\to s ll$ and $b\to s \nu\bar{\nu}$ observables. 
We found that $A_{CP} (K^{(*)})$ can be enhanced only in high-$q^2$ region up to $\sim -8\%$ for $K$-mode and up to $\sim -4\%$ for $K^*$-mode, however $R_{K^{(*)}}^{\nu\bar{\nu}}$ observables are no more than $1.35$.

Next, we observe that the triple product $A_7$, $A_8$, $A_9$ asymmetries are more prominent to the new CP violating phase, and can attain a few percent in the central- and high-$q^2$. However, $A_{CP}$ in the central-$q^2$ $\sim 0$, but in the high-$q^2$ can be enhanced up to $-4\%$.  Furthermore, these observables are more attractive from experimental point of view.

Therefore the observation of $A_{CP}$ as well as CP violating angular observables will not only provide an evidence of new physics with complex phase but their accurate measurements would also facilitate the unique identification of possible new physics in the decays induced by the $b\to s$ transition. The direct asymmetry can be measured at the LHCb or Belle-II \cite{Belle-II:2018jsg}, \cite{Lopes:2005km}. However, the measurement of the CP violating angular observables require higher statistics which can be attained at the HL-LHC in narrow bins of $q^2$  \cite{Cerri:2018ypt}. 

As for the future prospects of $A_i$, $S_i$  and $A_{CP}$ measurements for $B^0\to K^* \mu^+\mu^-$ decay (see, e.g.,   Ref.~\cite{LHCb:2015svh,LHCb:2020lmf,LHCb:2022ine}), we have the following situation\footnote{Note, that further uncertainty estimates will be assumed in absolute units. }. With 3$fb^{-1}$ \cite{LHCb:2015svh} LHCb measures CP asymmetries in $B\to K^{(*)} l^+l^-$ with $\sim 4-6\%$ uncertainties.  However, with increasing the collected luminosity up to 4.7$fb^{-1}$ \cite{LHCb:2020lmf} the uncertainties are estimated to be $\sim 2-4\%$. During LHC Runs 3 and 4, with a goal to collect 50$fb^{-1}$ of data \cite{LHCb:2022ine}, the statistical uncertainties can be potentially decreased to $\sim 1-1.5\%$. At the end, further Upgrades called Ib and II planning to collect 300$fb^{-1}$ \cite{LHCb:2022ine}. In this case the statistical uncertainties are $\sim 0.4-0.6\%$, which is near the current systematical one. Thus, the enhancements in $A_8$ and $A_{CP}(K)$ predicted by FIT$_2$ can be tested experimentally.

The  dineutrino modes can be studied by the Belle II experiment. According to Ref.~\cite{Belle-II:2022cgf}, with 50$ab^{-1}$ of data the uncertainties on the signal strengths with respect to the SM (corresponding to $R^{\nu\bar\nu}_K$ and $R^{\nu\bar\nu}_{K^*}$) can reach 0.08 ($K^+$) and 0.23 ($K^{*0}$). Obviously, this is not enough to favour or exclude our benchmark points. Nevertheless, as seen from Fig.~\ref{fig:RKnu_ACP}, some scenarios lying in the vicinity of the FIT$_2$, predict $R^{\nu\bar\nu}_K\sim1.3-1.35$, and, thus,  can be probed by future Belle II measurements.

\section{Acknowledgement }

Financial support from the Grant of the Russian Federation Government, Agreement No. 14.W03.31.0026 from 15.02.2018 is kindly acknowledge.

\appendix

\section{Model predictions for CP-averaged angular observables\label{sec:CPaveraged}}

In the following Tables \ref{tab:pred_S3456} and \ref{tab:pred_S789} we give  our predictions for the $S_i$ observables averaged over central- and high-$q^2$ bins. The SM value and experimental results \cite{LHCb:2020lmf} are indicated. FIT$_1$ corresponds to real parameters, while FIT$_2$ takes into account two NP quark phases. 

\begin{table}[h]
\centering
\caption{Prediction of various CP-averaged angular observables  in $B^0\to K^* \mu^+\mu^-$ in the central- and high-$q^2$ region.}
\label{tab:pred_S3456}
 \begin{tabular}{|c|c|c|c|c|}
			\hline
		 & $S_3^{[1.1,6]}$(\%)& $S_4^{[1.1,6]}$(\%)& $S_5^{[1.1,6]}$(\%)&$A_{FB}^{[1.1,6]}$(\%)\\ 
		    \hline
      SM  \cite{Straub:2018kue}&$-1.31 \pm 0.55$&$-14.8\pm 2.0$&$-18.6\pm 3.8$&$0.9\pm 2.9$\\
      \hline 
		    EXP \cite{LHCb:2020lmf}&$-1.2\pm 2.5\pm 0.3$&$-13.6 \pm 3.9\pm 0.3$&$-5.2 \pm 3.4\pm 0.7$&$-7.3\pm 2.1\pm 0.2$\\
		\hline
		FIT$_1$ &$-0.94 \pm 0.64$&$-14.8 \pm 2.15$&$-8.1 \pm 4.56$&$-5.79 \pm 3.86$\\
		\hline
		FIT$_2$ &$-1.2 \pm 0.65$&$-15.0 \pm 2.10$&$-8.2 \pm 4.60$&$-5.61 \pm 3.91$\\
		\hline\hline
	& $S_3^{[15,19]}$(\%)& $S_4^{[15,19]}$(\%)& $S_5^{[15,19]}$(\%)&$A_{FB}^{[15,19]}$(\%)\\ 
		    \hline
      SM  \cite{Straub:2018kue}&$-20.5 \pm 2.1$&$-30.0\pm 0.8$&$-28.0\pm 2.2$&$36.8\pm 2.7$\\
      \hline
		     EXP \cite{LHCb:2020lmf}&$-18.9 \pm 3.0\pm 0.9$&$-30.3 \pm 2.4\pm 0.8$&$-31.7 \pm 2.4\pm 0.11$&$35.3\pm2.0\pm 1.0$\\
		\hline
		FIT$_1$ &$-20.1 \pm 2.2$&$-30.2 \pm 0.85$&$-24.3 \pm 2.3$&$31.60 \pm 3.54$\\
		\hline
		FIT$_2$ &  $-20.6 \pm 2.2$&$-30.1 \pm 0.90$&$-24.3 \pm 2.4$&$31.54 \pm 3.60$\\
		\hline
		\end{tabular}
\end{table}	
\begin{table}[ht]
\centering
\caption{Prediction of various CP-averaged angular observables  in $B^0\to K^* \mu^+\mu^-$ in the central- and high-$q^2$ region.}
\label{tab:pred_S789}
 \begin{tabular}{|c|c|c|c|c|c|c|c|}
			\hline
		 &  $S_7^{[1.1,6]}$(\%)&$S_8^{[1.1,6]}$(\%) &$S_9^{[1.1,6]}$(\%)&$F_L^{[1.1,6]}$\\
   \hline
      SM \cite{Straub:2018kue}&$-1.9 \pm 3.9$&$-0.6 \pm 1.5$&$-0.07 \pm 0.24$&$0.750\pm 0.044$\\
		    \hline
		    EXP \cite{LHCb:2020lmf}& $-9.0\pm 3.4\pm 0.2$& $-0.9 \pm 3.7\pm 0.2$&$-2.5\pm 2.6\pm 0.2$&$0.700 \pm 0.025\pm 0.013$\\
		\hline
		FIT$_1$ &$-2.12 \pm 3.95$&$-0.46 \pm 1.54$&$-0.10 \pm 0.30$&$0.710 \pm 0.06$\\
		\hline
		FIT$_2$ &$-2.10\pm 4.00$&$-0.49\pm 1.57$ &$-0.08 \pm 0.32$&$0.708 \pm 0.07$\\
		\hline\hline
	& $S_7^{[15,19]}$(\%)&$S_8^{[15,19]}$(\%) &$S_9^{[15,19]}$(\%)&$F_L^{[15,19]}$\\
 \hline
      SM \cite{Straub:2018kue}&$-0.10 \pm 3.20$&$0.02 \pm 0.95$&$0.02 \pm 1.10$&$0.340\pm 0.029$\\
		    \hline
		     EXP \cite{LHCb:2020lmf}& $3.5\pm 3.0\pm 0.3$& $0.5 \pm 3.1\pm 0.1$&$-3.1\pm 2.9\pm 0.1$&$0.345\pm 0.020\pm 0.007$\\
		\hline
		FIT$_1$ &$-0.12 \pm 3.21$&$0.14 \pm 0.97$&$0.21 \pm 1.13$&$0.342 \pm 0.03$\\
		\hline
		FIT$_2$ &  $-0.12 \pm 3.24$&$0.08\pm 0.98$&$0.11 \pm 1.15$&$0.340 \pm 0.03$\\
		\hline
		\end{tabular}
\end{table}

\bibliography{references.bib}
\end{document}